\documentclass[american,aps,prd,10pt,nofootinbib,superscriptaddress,preprintnumbers]{revtex4-2}
\usepackage[T1]{fontenc}
\usepackage[latin9]{inputenc}
\usepackage{xcolor}
\usepackage{amsmath}
\usepackage{amssymb}
\usepackage{graphicx}
\usepackage{wasysym}

\makeatletter
\def\Mpl{M_{\rm P}}

\makeatother

\usepackage{babel}
\begin{document}
\preprint{YITP-22-142, IPMU22-0062}
\title{Gravitational collapse and formation of a black hole in a type II minimally modified gravity theory}
\author{Antonio De Felice}
\affiliation{Center for Gravitational Physics and Quantum Information, Yukawa Institute
for Theoretical Physics, Kyoto University, 606-8502, Kyoto, Japan}
\author{Kei-ichi Maeda}
\affiliation{Department of Physics, Waseda University, Shinjuku, Tokyo 169-8555,
Japan}
\affiliation{Center for Gravitational Physics and Quantum Information, Yukawa Institute
for Theoretical Physics, Kyoto University, 606-8502, Kyoto, Japan}
\author{Shinji Mukohyama}
\affiliation{Center for Gravitational Physics and Quantum Information, Yukawa Institute
for Theoretical Physics, Kyoto University, 606-8502, Kyoto, Japan}
\affiliation{Kavli Institute for the Physics and Mathematics of the Universe (WPI),
The University of Tokyo, Kashiwa, Chiba 277-8583, Japan}
\author{Masroor C.\ Pookkillath}
\affiliation{Center for Gravitational Physics and Quantum Information, Yukawa Institute
  for Theoretical Physics, Kyoto University, 606-8502, Kyoto, Japan}
\affiliation{Centre for Theoretical Physics and Natural Philosophy, Mahidol University, Nakhonsawan Campus,  Phayuha Khiri, Nakhonsawan 60130, Thailand}

\date{\today}
\begin{abstract}
We study the spherically symmetric collapse of a cloud of dust in VCDM, a class of gravitational theories with two local physical degrees of freedom. We find that the collapse corresponds to a particular foliation of the Oppenheimer-Snyder solution in general relativity (GR) which is endowed with a constant trace for the extrinsic curvature relative to the time $t$ constant foliation. For this solution, we find that the final state of the collapse leads to a static configuration with the lapse function vanishing at a radius inside the apparent horizon. Such a point is reached in an infinite time-$t$ interval, $t$ being the cosmological time, i.e.\ the time of an observer located far away from the collapsing cloud. The presence of this vanishing lapse endpoint implies the necessity of a UV completion to describe the physics inside the resulting black hole. On the other hand, since the corresponding cosmic time $t$ is infinite, VCDM can safely describe the whole history of the universe at large scales without knowledge of the unknown UV completion, despite the presence of the so-called shadowy mode. 
\end{abstract}
\maketitle

\section{Introduction}

It is a time in cosmology when we are awaiting for some answers to
fundamental questions. What is gravity? The answer to this question
might be still too far in time or even too difficult to be understood.
The answer might in fact connect the quantum realm to the large scales
of the universe. After all, even before aiming to find an answer to
this question, currently we need to understand why cosmological data today
seem to give a puzzling picture of the theory which is needed to model
them.

In fact, on assuming all experiments being free from significant
unknown systematics, it seems impossible to fit all the data at hand
by means of $\Lambda$CDM.  Therefore we are bound to explore
alternative theories which may be giving instead a better fit to the
data, fixing current cosmological tensions. In order to address this
point, namely to have a theory which would allow general (but non
vanishing) $H(z)$ and even a general (but positive) $G_{{\rm
    eff}}(z)/G_{N}$, the theories of VCDM and VCCDM were introduced
\citep{DeFelice:2020eju,DeFelice:2020prd}. Here, by $G_{\rm eff}$ we mean the
effective gravitational constant appearing in the modified Poisson equation which relates the over density perturbation of
the matter, $\delta_{\rm m}$, to the Bardeen gravitational potential
$\Psi$.  On top of these interesting phenomenological properties, both
VCDM and VCCDM only possess two degrees of freedom in the gravity
sector, i.e.\ the two polarization modes of the tensorial gravitational
waves. These two theories coincide with each other whenever cold dark
matter components are negligible or not considered.

In the high $k$ limit the VCDM theory acts like GR, i.e., $G_{\rm
  eff}/G_{N} \rightarrow 1$ (refer to~\cite{DeFelice:2020eju}). Due to
this fact VCDM theory cannot change $G_{{\rm eff}}(z)/G_{N}$ from
unity, while VCCDM can change $G_{cc}/G_{N}$, where $G_{cc}$ is the
self gravitational coupling between cold dark matter
particles. Therefore, VCDM and VCCDM will share the same solutions in
the vacuum case (e.g.\ same BH solutions~\cite{DeFelice:2022uxv}) as
well as solutions which only deal with baryonic matter (e.g.\ same
star solutions~\cite{DeFelice:2021xps}).

In both theories, we may find an appropriate cosmological model, which
explains the observational
data~\cite{DeFelice:2020prd,DeFelice:2020cpt}. In addition to it, when
we focus on strongly gravitating compact objects, we have to reanalyze
their viability. As we have shown in the previous papers, VCDM and
VCCDM admit the Schwarzschild solution~\cite{DeFelice:2020onz,
  DeFelice:2022uxv} as well as a solution of TOV equation that is the
same as the one in GR~\cite{DeFelice:2021xps}.  In VCDM/VCCDM,
however, solutions with the same spacetime geometry but with different
time slicings are physically different since the theory has a
preferred time slicing. In order for a solution to be trustable, not
only the spacetime geometry but also the time slicing should be
regular. For example, the standard Schwarzschild-type foliation of a
spherically-symmetric, static metric is ill-defined at the horizon and
thus the corresponding solution in VCDM/VCCDM is physically singular
while in GR it is simply a coordinate singularity. Then the question
is how a black hole is formed in VCDM after gravitational collapse. In
the formation of a black hole, a singularity should not appear on or
outside the horizon. In this sense, any spherically-symmetric, static
solution with the standard Schwarzschild-type foliation is no longer
to be a valid black hole solution in VCDM after gravitational
collapse. Hence, we ask, what kind of black hole solution in VCDM is
found after gravitational collapse?

In this paper, in order to answer this question, we want to
investigate the spherically symmetric collapsing solutions (or, at
least, a subset of them) for a cloud of dust in VCDM (or for a cloud
of baryonic dust in VCCDM). In a recent paper of
ours~\cite{DeFelice:2022uxv}, we have found that there is a subset of
the solutions of VCDM theory which are also solutions of GR provided
that the trace of the extrinsic curvature $K$ (relative to the
$t$-constant hypersurface) is a constant and that the VCDM auxiliary
field $\phi$ is a constant as well.  Motivated by the aforementioned
result, we study the Oppenheimer-Snyder (OS) solution of GR rewritten
in a coordinate patch which allows $K=K_{\infty}$, and study the
matching conditions at the surface of the star for the metric field
and all the auxiliary fields of VCDM.

VCDM and VCCDM are classified as Type-IIa Minimally Modified Gravity
(MMG) theories~\cite{Aoki:2021zuy,DeFelice:2022uxv}. There are also
other minimally modified gravity theories, each of which propagates
only two local physical degrees of freedom~\cite{Afshordi:2006ad,
Maeda:2022ozc, Kohri:2022vst, DeFelice:2015hla, DeFelice:2015moy,
DeFelice:2021trp, DeFelice:2020prd, DeFelice:2017wel,
DeFelice:2017rli, Mukohyama:2019unx, Aoki:2020oqc, Aoki:2020lig,
DeFelice:2020ecp, DeFelice:2022mcd}. They break general four
dimensional diffeomorphism invariance, because of the presence of a
shadowy mode which defines a preferred foliation. For this preferred
foliation the shadowy mode satisfies an elliptic equation of
motion. Therefore different foliations mean physically different
solutions in VCDM. For example in cosmology the trace of the extrinsic
curvature $K$ is a function of time and thus the theory is different
from $\Lambda$CDM and has interesting
phenomenology~\cite{DeFelice:2020cpt}. On the other hand, in the case
of black holes, if we set $K$ and $\phi$ to be constants then VCDM
admits GR solutions. As a consequence, not all the slicings of the
GR-OS solution will be appropriate (i.e.\ allowed by the equations of
motion) slicings in VCDM. In particular the standard slicing of the OS
solution provides an interior solution which is homogeneous and
isotropic. In this case one finds the trace of the extrinsic curvature
is merely proportional to the contraction rate $H(t)$ for the interior
solution. But this means that at least this foliation cannot provide a
solution of GR and that of VCDM at the same time. So one then wonders
how a collapse in VCDM looks like.

In this paper we indeed show that a different foliation of the OS
solution can describe a solution for both VCDM and GR. On following
the results of \citep{DeFelice:2022uxv}, this VCDM-GR compatible
foliation is defined as to make the trace of the extrinsic curvature
for the solution a constant. This constant, after imposing the Israel
junction conditions, needs to correspond to the trace of the extrinsic
curvature for the outer metric, which then matches a constant de Sitter
solution at infinity.

This leads to an interesting collapse solution in VCDM which then
predicts the final state of the solution as reaching a radius (located
inside the apparent horizon) where the lapse function vanishes (there
and at any place up to the origin). This endpoint is reached in an
infinite $t$-time interval, $t$ being the time of a cosmological
observer, i.e.\ an observer located far away from the collapsing
cloud. The solutions, in this limit, reduces to the static solutions
found in \citep{DeFelice:2018vza}, with the difference that the free
parameters of the solutions are fixed by the collapse dynamics.

This paper is organized as following. At first, we introduce the VCDM
covariant action in Section~\ref{sec:vcdm-action}. Then, in Section
\ref{sec:The-outer-metric} we rewrite the GR (and VCDM) solutions in
vacuum, corresponding to the outer metric patch, having the property
that $K=K_{\infty}={\rm constant}$. We find the properties of the collapse
for the outer solutions in Section
\ref{sec:Collapsing-outside-vacuum}.  Instead, we study the interior
solution and its collapse in Section
\ref{sec:Collapsing-inside-dust}. We set the Israel junction
conditions for VCDM in Section \ref{sec:Matching-conditions}. On
fixing the appropriate boundary conditions at infinity, we finally
study the collapse and its final point, which then leads
asymptotically to static VCDM/GR solutions in Section
\ref{sec:Analytical-and-numerical}.  We finally give our conclusions
in Section \ref{sec:Discussion}.

\section{The VCDM action\label{sec:vcdm-action}}

Here we show covariant action for the VCDM theory, which was first
introduced in~\cite{DeFelice:2022uxv}:
\begin{eqnarray}
S & = & \Mpl^{2}\int d^{4}x\sqrt{-g}\left\{ \tfrac{1}{2}\,R^{(4)}-V(\phi)
-\tfrac{3}{4}\,\lambda^{2}-\lambda\,(\nabla^{\sigma}\mathfrak{n}_{\sigma}+\phi)\right.\nonumber \\
 & + & \left.\frac{\lambda_{2}}{\cal A}\,[\gamma^{\tau\rho}\nabla_{\tau}\nabla_{\rho}\phi
 +\mathfrak{n}^{\rho}(\nabla_{\rho}\phi)\,\nabla^{\sigma}\mathfrak{n}_{\sigma}]+\lambda_{T}\,(1
 +g^{\mu\nu}\mathfrak{n}_{\mu}\mathfrak{n}_{\nu})\right\} ,\label{eq:cov_act}\\
\mathfrak{n}_{\mu} & \equiv & -{\cal A}\,\nabla_{\mu}\mathcal{T}\,,\\
\gamma^{\mu\nu} & = & g^{\mu\nu}+\mathfrak{n}^{\mu}\mathfrak{n}^{\nu}\,,\label{vcdm-action1}
\end{eqnarray}
where $\phi$, $\cal A$ and $\mathcal{T}$ are 4D scalars. We have
Lagrange multipliers $\lambda$, $\lambda_{2}$ and $\lambda_{T}$, which
will determine the connections between the 4D scalars and the
geometrical objects. Choosing the unitary gauge for the time
coordinate, this action reduces to the equivalent action which was introduced
in~\cite{DeFelice:2020eju}.

We can integrate out the field $\mathcal{A}$ by varying $\lambda_{T}$, and the action can then be rewritten as
\begin{eqnarray}
S & = & \Mpl^{2}\int d^{4}x\sqrt{-g}\left\{ \tfrac{1}{2}\,R^{(4)}-V(\phi)-\tfrac{3}{4}\,\lambda^{2}
-\lambda\,(\nabla^{\sigma}\mathfrak{n}_{\sigma}+\phi)\right.\nonumber \\
 & + & \left.(-g^{\mu\nu}\nabla_{\mu}\mathcal{T}\,\nabla_{\nu}\mathcal{T})^{1/2}\,\lambda_{2}\,[\gamma^{\tau\rho}\nabla_{\tau}\nabla_{\rho}\phi
 +\mathfrak{n}^{\rho}(\nabla_{\rho}\phi)\,\nabla^{\sigma}\mathfrak{n}_{\sigma}]\right\} ,\label{eq:covariant_vcdm_action}\\
\mathfrak{n}_{\mu}& = & -(-g^{\mu\nu}\nabla_{\mu}\mathcal{T}\,\nabla_{\nu}\mathcal{T})^{-1/2}\,\nabla_{\mu}\mathcal{T}\,,\\
\gamma^{\mu\nu} & = & g^{\mu\nu}+\mathfrak{n}^{\mu}\mathfrak{n}^{\nu}\,,\label{vcdm-action2}
\end{eqnarray}
so that $\nabla_{\mu}\mathcal{T}$ is timelike by construction. From this action we can find the covariant equations of motion, which were derived in Appendix A of~\cite{DeFelice:2022uxv} (with a slightly different notation).

In what follows, using this VCDM model, we look for a spherically
symmetric time-dependent solution which describes the gravitational dust
collapse and the formation of a black hole. Note that the presented
solution can also be applied to the VCCDM model.

\section{The outer metric\label{sec:The-outer-metric}}

In this section we write the Schwarzschild-de Sitter metric, i.e.\ the
vacuum spherically symmetric GR solutions in the presence of a
cosmological constant. It is given in such a way that the extrinsic
curvature of the $t$-constant hypersurface has a constant trace. Then, as found in
\citep{DeFelice:2022uxv}, automatically this solution will be also a
solution of VCDM, provided that also the auxiliary field $\phi$ is constant.

\subsection{Schwarzschild reloaded}

Let us then assume an outer metric with the following ansatz
\begin{eqnarray}
ds^{2} & = & -\alpha(t,r)^{2}\,[1-\beta(t,r)^{2}/F(t,r)]\,\dot{T}^{2}\,dt^{2}+2\,\frac{\alpha\beta}{F}\,\dot{T}\,dt\,dr+\frac{dr^{2}}{F}+r^{2}d\Omega^{2}\,,\label{eq:outer_metric}\\
d\Omega^{2} & = & \frac{dz^{2}}{1-z^{2}}+(1-z^{2})\,d\theta_{2}^{2}\,,
\end{eqnarray}
where we have defined $z=\cos\theta_{1}$, and have introduced a function
$\dot{T}>0$, explicitly showing time reparametrization invariance,
which is a symmetry of VCDM. Then for this spherically symmetric ansatz
we will consider the $t$-extrinsic curvature, that is the extrinsic
curvature of the hypersurface $t={\rm const.}$ with normal vector $\mathfrak{n}_{\alpha}dx^{\alpha}=-\alpha\dot{T}dt$,
with $\mathfrak{n}_{\alpha}\mathfrak{n}_{\beta}g^{\alpha\beta}=-1$,
as 
\begin{equation}
-K=\frac{2\beta}{r}+\frac{\beta\alpha_{,r}}{\alpha}+\beta_{,r}-\frac{\beta F_{,r}}{2F}+\frac{F_{,t}}{2F\alpha\dot{T}}\,.
\end{equation}
We want to have a foliation of the spacetime, in which the trace of the extrinsic curvature takes the value
\begin{equation}
K=K_{\infty}={\rm constant}\,,\label{eq:K_const}
\end{equation}
so that this solution belongs to the class of VCDM solutions which
are also solutions of GR (provided that $\phi$ is also a constant).

We can formally solve the PDE in Eq.\ (\ref{eq:K_const}), in terms
of $\alpha$, by finding
\begin{equation}
\alpha=\left\{W(t)-\frac{1}{\dot{T}}\int^{r}\frac{F_{,t}
}{2F\beta}\exp
\left[
{-\int^{r_{2}}\left(-\frac{K_{\infty}}{\beta}+\frac{F_{,r}}{2F}-\frac{\beta_{,r}}{\beta}-\frac{2}{r}\right){\rm d}r_{1}}\right] {\rm d}r_{2}
\right\}\exp\left[{\int^{r}\left(-\frac{K_{\infty}}{\beta}+\frac{F_{,r}}{2F}-\frac{\beta_{,r}}{\beta}-\frac{2}{r}\right){\rm d}r_{3}}
\right]\,.\label{alpha_sol}
\end{equation}

Given the metric in Eq.\ (\ref{eq:outer_metric}), we can find its
own Einstein tensor, $G^{\mu}{}_{\nu}$. In fact, its $(t,r)$ component
can be found to be
\begin{equation}
G^{t}{}_{r}=-(K_{\infty}r+r\beta_{,r}+2\beta)\,[\dots]\,,
\end{equation}
where the dots represent, in general, a non-zero quantity (which,
if vanishing, would make $\alpha$ not well defined). Setting this
Einstein tensor component, $G^{t}{}_{r}$, to vanish leads to
\begin{equation}
\beta=-\frac{1}{3}\,K_{\infty}r+\frac{\kappa(t)}{r^{2}}\,.
\end{equation}
With the above solution for $\beta$ we find the following relation
\begin{equation}
\exp\left[{\int^{r}\left(-\frac{K_{\infty}}{\beta}+\frac{F_{,r}}{2F}-\frac{\beta_{,r}}{\beta}-\frac{2}{r}\right){\rm d}r_{1}}\right]=
\exp\left[{\int^{r}\left(\frac{F_{,r}}{2F}\right){\rm d}r_{1}}\right]=W_{1}(t)\,\sqrt{F}\,,
\end{equation}
and then Eq.\ (\ref{alpha_sol}) can be written as
\begin{eqnarray}
\alpha & = & \left(W(t)W_{1}(t)-\frac{W_{1}(t)}{\dot{T}W_{2}(t)}\int^{r}\frac{F_{,t}}{2F\beta\sqrt{F}}{\rm d}r_{1}\right)\,\sqrt{F}\nonumber \\
 & = & \left(1-\frac{1}{\dot{T}}\int^{r}\frac{F_{,t}}{2\beta F^{3/2}}{\rm d}r_{1}\right)\,\sqrt{F}\,,
\end{eqnarray}
having fixed the integration functions, as to make $\alpha\to1$ for $F_{,t}\to0$ when $F\to1$.

In cosmology in GR, on de Sitter with the standard flat chart
coordinates, we have $K=K_{\infty}=3H_{\infty}$. In this case the
first Friedmann equation leads to
$3\Mpl^{2}H_{\infty}^{2}=\Mpl^{2}\Lambda_{{\rm eff}}$, or
$\Lambda_{{\rm
    eff}}=3H_{\infty}^{2}=\frac{1}{3}\,K_{\infty}^{2}$. Then this
fixes $K_{\infty}$ in terms of the cosmological constant, or in terms
of $H_{\infty}$, for this foliation. Therefore let us now fix the
Schwarzschild background to satisfy the Einstein equation as
\begin{equation}
G^{\mu}{}_{\nu}=-\Lambda_{{\rm
    eff}}\delta^{\mu}{}_{\nu}=-\frac{1}{3}\,K_{\infty}^{2}\,\delta^{\mu}{}_{\nu}.
\end{equation}

If we solve the time-time component, that is $G^{t}{}_{t}=-\frac{1}{3}\,K_{\infty}^{2}$, then we find
\begin{equation}
F=1+\frac{Z(t)}{r}+\frac{\kappa(t)^{2}}{r^{4}}\,.
\end{equation}
From the $r$-$r$ component, that is
$G^{r}{}_{r}=-\frac{1}{3}\,K_{\infty}^{2}$, we need to set the following
constraint
\begin{equation}
(-2K_{\infty}\dot{\kappa}-3\dot{Z})\,[\dots]=0\,,
\end{equation}
where, once more, the dots represent, in general, a non-zero quantity.
Then we can solve
\begin{equation}
Z(t)=-r_{H}-\frac{2}{3}\,K_{\infty}\,\kappa(t)\,,
\end{equation}
where $r_{H}$ is an integration constant. Then finally
\begin{align}
\label{eq:F}
F & =1-\frac{r_{H}}{r}-\frac{2}{3}\frac{K_{\infty}\kappa(t)}{r}+\frac{\kappa(t)^{2}}{r^{4}}\,,\\
\beta & =-\frac{1}{3}\,K_{\infty}r+\frac{\kappa(t)}{r^{2}}\,,
\label{eq:Beta}
\end{align}
so that
\begin{equation}
F_{,t}=\frac{2\dot{\kappa}}{r^{2}}\,\beta\,,
\end{equation}
and
\begin{equation}
\alpha=\left(1-\frac{\dot{\kappa}}{\dot{T}}\int_{r_{0}}^{r}\frac{1}{r_{1}^{2}F^{3/2}}{\rm d}r_{1}\right)\,\sqrt{F}\,,
\end{equation}
where $r_0$ is just an integration constant.
The other equations of motion are also satisfied, so that the solution
is determined (up to time reparametrization $\dot{T}$) by the free
function $\kappa(t)$, and the constants $K_{\infty}$, and $r_{0}$.
In particular, we will set $r_{0}\to\infty$ as to have $\lim_{r\to\infty}\alpha=1$.

In summary, this is a solution which is nothing but the standard
Schwarzschild de Sitter of GR in a constant mean curvature slicing,
having a constant trace for the $t$-extrinsic curvature $K$, making
this solution also a solution of VCDM~\cite{DeFelice:2022uxv}
(provided also $\phi$ is a constant).  Since this solution is also a
solution of GR, it is not a surprise that this solution were already
found in the literature in the context of maximal slicing or constant
mean curvature slicing (see e.g.\
\citep{Estabrook:1973ue,Eardley:1979,Maeda:1980,Petrich:1985,Nakao:1990gw}.)
Note that this solution is a subset of the general time-dependent
solutions found in VCDM. Indeed, there exist other time-dependent VCDM
solutions which are not found in GR. However we study just this
GR-type solution in this paper since we would like to show the
existence of VCDM solution which describes gravitational collapse and
formation of a black hole.

Although this solution holds true for any value of $K_{\infty}$, we need
to set $K_{\infty}=3H_{\infty}=\frac{3}{2}\,V_{,\phi_{\infty}}-\phi_{\infty}$, the last
relation holding true for VCDM, as to have a de Sitter limit at infinity\footnote{Outside the de Sitter limit, it is not trivial to find a solution which can be extended up to infinity. This because in a universe with matter, before reaching $r$-infinity, matter sources (baryons, dark matter) will give non negligible contributions (sourcing furthermore a time dependence for the field $\phi$) which lay outside the validity of these $\Lambda_{\rm eff}$-vacuum solutions.}.
Then its contribution, corresponding to an effective cosmological
constant contribution, on astrophysical scales/configurations,
can be safely neglected, so that when we deal with numerics we will
set $K_{\infty}$ to vanish.

\subsection{Presence of the apparent horizons}

We have just seen that a constant trace of the extrinsic curvature
$K$, relative to the $t$-slicing hypersurface, corresponds to having
\begin{equation}
F(t,r)=1-\frac{r_{H}}{r}-\frac{2}{3}\frac{K_{\infty}\kappa(t)}{r}+\frac{\kappa(t)^{2}}{r^{4}}\,.
\end{equation}
In the above expression we have a free time-dependent function
$\kappa(t)$.  Let us see how the freedom of $\kappa(t)$ affects (or
not) the presence of the apparent horizon for the metric in
Eq.\ (\ref{eq:outer_metric}).  For this, once more we consider a
completely general spherically symmetric ansatz rewritten for our
convenience as
\begin{equation}
ds^{2}=-(N^{2}-B^{2})\,dt^{2}+2B\mathcal{F}dt\,dr+\mathcal{F}^{2}\,dr^{2}+r^{2}d\Omega^{2}\,.\label{eq:outer_metric_2}
\end{equation}
In the following we consider $N>0$, as well as $\mathcal{F}$ and $B$
(a negative $B$ would be regarded as a flip in $t\to-t$), and all of
them are functions of $t$ and $r$. Then let us consider two lightlike
vectors
\begin{eqnarray}
l^{\mu}\partial_{\mu} & = & \partial_{t}+\frac{N-B}{\mathcal{F}}\,\partial_{r}\,,\\
w^{\mu}\partial_{\mu} & = & \partial_{t}-\frac{N+B}{\mathcal{F}}\,\partial_{r}\,,
\end{eqnarray}
which satisfy
\begin{equation}
l^{\mu}l_{\mu}=0=w^{\mu}w_{\mu}\,,\qquad l^{\mu}w_{\mu}=-2N^{2}\,<0.
\end{equation}
Using the above lightlike vectors, we can define an orthogonal
projector
\begin{equation}
h_{{\rm lw}}^{\mu\nu}=g^{\mu\nu}+\frac{l^{\mu}w^{\nu}+l^{\nu}w^{\mu}}{(-l^{\sigma}w_{\sigma})}\,,
\end{equation}
so that $h_{{\rm lw}}^{\mu\nu}l_{\nu}=0=h_{{\rm lw}}^{\mu\nu}w_{\nu}$.
Then we can build the following two scalars
\begin{equation}
\theta_{l}=h_{{\rm lw}}^{\mu\nu}\nabla_{\nu}l_{\mu}=\frac{2}{r\mathcal{F}}\,(N-B)\,,\qquad\theta_{w}=h_{{\rm lw}}^{\mu\nu}\nabla_{\nu}w_{\mu}=-\frac{2}{r\mathcal{F}}\,(N+B)\,.
\end{equation}
Let us now consider the marginally outer trapped surface (MOTS), i.e.\ the
apparent horizon, that is the surface $r=r_{AH}(t)$ for which 
\begin{equation}
\theta_{l}(t,r=r_{AH}(t))=0\,,\qquad\theta_{w}(t,r=r_{AH}(t))<0\,.
\end{equation}
In this case we have a non trivial solution of the form
\begin{equation}
N(t,r_{AH}(t))=B(t,r_{AH}(t))\,,\qquad\textrm{for which}\qquad g^{rr}(t,r=r_{AH}(t))=\frac{N^{2}-B^{2}}{N^{2}F^{2}}=0\,.
\end{equation}
So we have a MOTS, an apparent horizon at $r=r_{AH}(t)$. Let us see
what happens for the metric under study. On identifying the metric
in Eq.\ (\ref{eq:outer_metric}) with the one in Eq.\ (\ref{eq:outer_metric_2}),
we have
\begin{eqnarray}
\mathcal{F}^{2} & = & \frac{1}{F}\,,\qquad{\rm or}\qquad\mathcal{F}=F^{-1/2}\,,\\
B\mathcal{F} & = & \frac{\alpha\beta}{F}\,\dot{T}\,,\qquad{\rm or}\qquad B=\frac{\alpha\beta\dot{T}}{\sqrt{F}}\,,\\
N^{2}-B^{2} & = & \alpha^{2}\,\dot{T}^{2}\,[1-\beta^{2}/F]\,,\qquad{\rm or}\qquad N=\alpha\,\dot{T}\,.
\end{eqnarray}
In this case we have the MOTS for a nonzero $\alpha$ and $\dot{T}$, if
\begin{equation}
\alpha\dot{T}=\frac{\alpha\beta\dot{T}}{\sqrt{F}}\,,\qquad{\rm or}\qquad F=\beta^{2}\,.
\end{equation}
The above equation on using Eq.~(\ref{eq:F}) and Eq.~(\ref{eq:Beta}) leads to
\begin{equation}
1-\frac{r_{H}}{r_{AH}}-\frac{2}{3}\frac{K_{\infty}\kappa(t)}{r_{AH}}+\frac{\kappa(t)^{2}}{r_{AH}^{4}}=\frac{1}{9}\,K_{\infty}^{2}r_{AH}^{2}+\frac{\kappa(t)^{2}}{r_{AH}^{4}}-\frac{2}{3}\,\frac{K_{\infty}\kappa(t)}{r_{AH}}
\end{equation}
or
\begin{equation}
1-\frac{r_{H}}{r_{AH}}-\frac{1}{9}\,K_{\infty}^{2}r_{AH}^{2}=0\,.
\end{equation}
This shows that there are only two horizons, the event horizon and the
cosmological horizon, and also that $\dot{r}_{AH}=0$, independently of
the function $\kappa(t)$. So the free function $\kappa(t)$ does not
affect the position/behavior of the apparent horizon and we are still
able to choose it the way we think it is more useful, at least
locally.  However, as we shall see later on, the value of $\kappa(t)$
will be determined by the dynamics of the collapse.

\subsection{VCDM coordinate patch}

We know that an exact GR configuration/solution is also shared by the
VCDM theory, provided that the VCDM auxiliary field $\phi$ has a
constant profile and that the trace of the extrinsic curvature tensor
(for the $t$-foliation in VCDM) is also a constant. Then the previous
GR solution can be embedded also in VCDM, since $K=K_{\infty}$. The
vacuum spacetime has an apparent horizon at the zeros of the function
$1-\frac{r_{H}}{r}-\frac{1}{9}\,K_{\infty}^{2}r^{2}$.

We want to show here that it is possible that, for some value of
$\kappa(t)$ at a given time, the lapse function $N=\alpha\dot{T}$, for
$\dot{T}>0$, never vanishes for any $r>0$. In particular, for a
constant $\kappa(t)=\kappa_{0}$, we have that $\alpha=\sqrt{F}$, so
that also $F$ never vanishes for this coordinate patch. In any case,
the metric would still be possessing an apparent horizon so that the
singularity, which may appear at the center $r=0$, is never naked (for
$r_{H}>0$ and $K_{\infty}^{2}r_{H}^{2}\ll1$).  For a non-vanishing
$F$, this coordinate choice would be able to describe all the $r>0$
region.

Hence, for simplicity, in the following we only consider a constant
$\kappa$ and look for the domain of $\kappa$'s which makes $F$ (or
$\alpha$) never vanish. The study of this coordinate patch will help
us understanding the final state of the collapse in VCDM theory. In
particular, let us try to set the equation $r^{4}F=0$, as to have
double real roots and two complex ones. Let us try to find the
``extremal'' value of $\kappa$ for which this happens.

Then in this case, on assuming $r>0$, let us find constants
$\xi_{1,2,3}$ to allow for this possibility to hold true. Then we have
\begin{equation}
r^{4}F=r^{4}-\left(r_{H}+\frac{2}{3}K_{\infty}\kappa\right)r^{3}+\kappa^{2}=\frac{1}{\xi_{1}}\,(r-r_{1})^{2}\,[\xi_{1}r^{2}+\xi_{2}r+\xi_{3}]\,,\label{eq:2r2c}
\end{equation}
with $\xi_{2}^{2}-4\xi_{1}\xi_{3}<0$ and $\xi_{1}\neq0$. The absence
of terms linear in $r$ in the rhs of Eq.\ (\ref{eq:2r2c}) imposes that
\begin{equation}
\xi_{3}=\frac{\xi_{2}r_{1}}{2}\,.
\end{equation}
Then the absence of quadratic terms in $r$ also imposes that
\begin{equation}
\xi_{2}=\frac{2\xi_{1}r_{1}}{3}\,,\qquad{\rm and}\qquad\xi_{3}=\frac{\xi_{1}r_{1}^{2}}{3}\,.
\end{equation}
At this point we have that
\begin{equation}
\xi_{2}^{2}-4\xi_{1}\xi_{3}=\left(\frac{2\xi_{1}r_{1}}{3}\right)^{2}-\frac{4\xi_{1}^{2}r_{1}^{2}}{3}=-2\,\left(\frac{2\xi_{1}r_{1}}{3}\right)^{2}<0\,,
\end{equation}
and
\begin{equation}
r^{4}-\left(r_{H}+\frac{2}{3}K_{\infty}\kappa\right)r^{3}+\kappa^{2}=r^{4}-\frac{4r_{1}r^{3}}{3}+\frac{r_{1}^{4}}{3}\,.
\end{equation}
For the last equation to make sense as an identity we need
\begin{align}
r_{1} & =\frac{1}{4}\left(3r_{H}+2\kappa K_{\infty}\right),\\
\kappa^{2} & =\frac{r_{1}^{4}}{3}=\frac{1}{3}\left[\frac{1}{4}\left(3r_{H}+2\kappa K_{\infty}\right)\right]^{4}.
\end{align}
From the first equation we can derive the value of $r_{1}$, and we can
see that for vanishing $K_{\infty}$ we indeed find $r_{1}=3r_{H}/4$,
as expected (see e.g.~\cite{DeFelice:2020onz}). The second equation
does not trivially hold, unless we find the values of $\kappa$ for
which the equation is satisfied. Therefore we find for positive
$\kappa$'s that
\begin{equation}
\kappa=\kappa_{+}\equiv\frac{9r_{H}^{2}}{-6K_{\infty}r_{H}+4\sqrt{3}\left(\sqrt{-2\sqrt{3}K_{\infty}r_{H}+4}+2\right)}=\frac{3}{16}\sqrt{3}r_{H}^{2}+\frac{9}{64}K_{\infty}r_{H}^{3}+\mathcal{O}[K_{\infty}^{2}r_{H}^{4}]\,,
\end{equation}
and for negative $\kappa$'s that
\begin{equation}
\kappa=\kappa_{-}=-\frac{9r_{H}^{2}}{4\sqrt{3}\left(\sqrt{4+2\sqrt{3}K_{\infty}r_{H}}+2\right)+6K_{\infty}r_{H}}=-\frac{3}{16}\sqrt{3}r_{H}^{2}+\frac{9}{64}K_{\infty}r_{H}^{3}+\mathcal{O}[K_{\infty}^{2}r_{H}^{4}]\,.
\end{equation}
Hence, we find the following two cases
(see the appendix for explicit calculations).
\begin{enumerate}
\item For $\kappa_{-}\leq\kappa<0$ or $0<\kappa\leq\kappa_{+}$, the
  coordinate patch can get inside the Schwarzschild radius but has a
  lower bound $r_1$, that is
\begin{align}
0<\kappa\leq\kappa_{+}&: \ r_{1,+}\leq r_1<r_{H}\,, \\
\kappa_{-}\leq\kappa<0&: \ r_{1,-}\leq r_1<r_{H}\,,
\end{align}
where 
\begin{align}
r_{1,+}&\equiv\frac{1}{4}\,(3r_{H}+2K_{\infty}\kappa_{+})=-\frac{\sqrt{3}\sqrt{-2\sqrt{3}K_{\infty}r_{H}+4}}{2K_{\infty}}+\frac{\sqrt{3}}{K_{\infty}}\approx\frac{3r_{H}}{4}+\frac{3\sqrt{3}}{32}K_{\infty}r_{H}^{2}\,,\\
r_{1,-}&\equiv\frac{1}{4}\,(3r_{H}+2K_{\infty}\kappa_{-})=-\frac{\sqrt{3}}{K_{\infty}}+\frac{\sqrt{3}\sqrt{4+2\sqrt{3}K_{\infty}r_{H}}}{2K_{\infty}}\approx\frac{3r_{H}}{4}-\frac{3\sqrt{3}}{32}K_{\infty}r_{H}^{2}\,,
\end{align}
and the numbers $r_{1,\pm}$ have been expanded in terms of the
small parameter $K_{\infty}r_{H}$. For $\kappa=0$, the solution cannot
get inside the Schwarzschild radius and the lapse vanishes at $r=r_{H}$.
At $r=r_{1}$, both the functions $F$ and $\alpha=\sqrt{F}$ vanish,
leading to a vanishing lapse function, in general.
\item For $\kappa<\kappa_{-}$ or $\kappa>\kappa_{+}$ the solutions of
$r^{4}F=0$ are all complex and $F$ never vanishes, so as $\alpha$.
\end{enumerate}

In any case, we stress once more that, although at this level, there
is the freedom of the choice of $\kappa$, we will see that at the
end of the collapse, $\kappa$ will be set to belong to the first
possibility, namely $\kappa\leq\kappa_{+}$ (for positive $\kappa$'s),
as to exclude the second case.

\section{Collapsing outside vacuum solution\label{sec:Collapsing-outside-vacuum}}

Based on the previous results, in the following we will consider the
outer vacuum solution endowed with a time-dependent $\kappa$. So
we will consider then as the outside solution (i.e.\ the solution
valid outside the collapsing star) the one given by

\begin{align}
F & =1-\frac{r_{H}}{r}-\frac{2}{3}\frac{K_{\infty}\kappa(t)}{r}+\frac{\kappa(t)^{2}}{r^{4}}\,,\\
\beta & =-\frac{1}{3}\,K_{\infty}r+\frac{\kappa(t)}{r^{2}}\,,\\
\alpha & =\left(1-\frac{\dot{\kappa}}{\dot{T}}\int_{r_{0}}^{r}\frac{1}{r_{1}^{2}F(t,r_{1})^{3/2}}\,{\rm d}r_{1}\right)\,\sqrt{F}\,,
\end{align}
and for the sake of clarity we rewrite the metric as 
\begin{align*}
ds^{2} & =-\frac{\alpha^{2}}{F}\,[F-\beta^{2}]\,\dot{T}^{2}\,dt^{2}+2\,\frac{\alpha\beta\dot{T}}{F}\,dt\,dr+\frac{dr^{2}}{F}+r^{2}d\Omega^{2}\,,\\
d\Omega^{2} & =\frac{dz^{2}}{1-z^{2}}+(1-z^{2})\,d\theta_{2}^{2}\,,
\end{align*}
where, once more, the function $\dot{T}(t)$ is due to time-reparametrization
invariance. In the following we will also define the functions $g_{1,2}$
as
\begin{eqnarray}
\alpha(t,r) & = & g_{2}(t,r)-\frac{\dot{\kappa}}{\dot{T}}\,g_{1}(t,r)\,,\\
g_{2}(t,r) & = & \sqrt{F}\,,\\
g_{1}(t,r) & = & \sqrt{F}\int_{r_{0}}^{r}\frac{{\rm d}r_{1}}{r_{1}^{2}[1-r_{H}/r-\frac{2}{3}\frac{K_{\infty}\kappa(t)}{r}+\kappa(t)^{2}/r^{4}]^{3/2}}\,.
\end{eqnarray}
This implicit form is convenient in order not to hide any contribution
coming from $\dot{T}$.

We will define the hypersurface for the external metric coordinates
on which we will match the two metrics by
\begin{equation}
\Phi(x^{\mu})=r-R(t)=0\,,
\end{equation}
where $R(t)$ is still a function of time to be determined, which
corresponds to the radius of the collapsing cloud/star as seen from
the outside metric point of view. Then the normal $n_{\mu}$ to this
surface is proportional to the gradient of $\Phi$, namely 
\begin{equation}
n_{\mu}dx^{\mu}\propto dr-\dot{R}\,dt\,,
\end{equation}
with the condition that $n_{\alpha}n^{\alpha}=1$ for the normal vector
to the hypersurface. Then we find
\begin{align}
n_{\mu}dx^{\mu} & =\frac{1}{\sqrt{\Delta_{n}}}\,(g_{2}\dot{T}-g_{1}\dot{\kappa})\,(dr-\dot{R}\,dt)\,,\\
\Delta_{n} & \equiv g_{2}^{2}\dot{T}^{2}\left(F-\beta^{2}\right)-2g_{2}\dot{T}[\dot{\kappa}(F-\beta^{2})\mathit{g}_{1}+\beta\dot{R}]+g_{1}^{2}\dot{\kappa}^{2}\left(F-\beta^{2}\right)+2\dot{R}g_{1}\beta\dot{\kappa}-\dot{R}^{2}\,.
\end{align}
Now we can define the following projection tensors
\begin{eqnarray}
h_{\alpha\beta} & = & g_{\alpha\beta}-n_{\alpha}n_{\beta}\,,\\
h_{\alpha}{}^{\beta} & = & h_{\alpha\mu}g^{\mu\beta}\,,
\end{eqnarray}
so that
\begin{equation}
h_{\alpha}{}^{\beta}n_{\beta}=0\,,\qquad h_{\alpha\beta}n^{\beta}=0.
\end{equation}
We can define then the extrinsic curvature for this hypersurface as
\begin{equation}
\mathcal{K}_{\mu\nu}=h_{\mu}{}^{\rho}h_{\nu}{}^{\sigma}n_{\rho;\sigma}\,.
\end{equation}
This extrinsic curvature tensor should not be confused with the extrinsic
curvature tensor $K_{\mu\nu}$ which was defined for the $t$-constant
hypersurfaces. We further define three vectors as
\begin{equation}
e^{\mu}{}_{a}=\frac{\partial x^{\mu}}{\partial y^{a}}\,,
\end{equation}
where $y^{a}$ are coordinates on the hypersurface, here chosen to
be ($t,z,\theta_{2}$). Then we have
\begin{equation}
e^{\mu}{}_{t}\,\partial_{\mu}=\partial_{t}+\dot{R}\,\partial_{r}\,,\qquad e^{\mu}{}_{z}\,\partial_{\mu}=\partial_{z}\,,\qquad e^{\mu}{}_{\theta_{2}}\,\partial_{\mu}=\partial_{\theta_{2}}\,,
\end{equation}
satisfying $n_{\mu}e^{\mu}{}_{a}=0$, out of which we can find twelve
scalars as
\begin{equation}
h_{ab}=h_{ba}=\left.h_{\mu\nu}e^{\mu}{}_{a}e^{\nu}{}_{b}\right|_{r=R(t)}\,,\qquad\mathcal{K}_{ab}=\mathcal{K}_{ba}=\left.\mathcal{K}_{\mu\nu}e^{\mu}{}_{a}e^{\nu}{}_{b}\right|_{r=R(t)}\,.
\end{equation}
We impose the standard Israel junction conditions at the surface of
the star, so that these expressions need to be continuous on the hypersurface
joining the two different GR solutions. Out of these scalars, given
the spherical symmetric ansatz, only the diagonal components do not
vanish.

\section{Collapsing inside dust solution\label{sec:Collapsing-inside-dust}}

Let us now study the collapse from the point of view of the inside
metric. We then write the inside metric as a spatially closed homogeneous and isotropic
metric, however, by choosing a quite general time slicing, as follows
\begin{equation}
ds^{2}=-a\bigl(f(t,\chi)\bigr)^{2}\,\left[f_{,t}dt+f_{,\chi}d\chi\right]^{2}+a\bigl(f(t,\chi)\bigr)^{2}\left[\frac{d\chi^{2}}{1-\chi^{2}}+\chi^{2}d\Omega^{2}\right],
\end{equation}
where we have started with the conformal time $\eta$ with $N(\eta)=a(\eta)$
and made the general coordinate transformation $\eta=f(t,\chi)$, that respects the spherical symmetry, to relate $\eta$ to the VCDM time $t$. The coordinate
transformation, i.e.\ the slicing is chosen as to have $K=K_{\infty}$
for the trace of the extrinsic curvature of the $t$-const hypersurface.
In fact we want the same VCDM slicing to hold both outside and inside
the hypersurface to accommodate the GR solution in VCDM, and for this purpose we need to require that $K$ for the $t={\rm constant}$ hypersurface be constant everywhere, so as the VCDM field $\phi$. In GR, although this choice is allowed, the standard coordinates
of the Oppenheimer-Snyder solution would be the simplest to implement.
In VCDM, however, different slicings correspond to physically different solutions,
so that we are looking for that particular slicing which satisfies
$K=K_{\infty}$. For this metric, we have that the normal to the $t={\rm constant}$
surface can be written as
\begin{equation}
\mathfrak{n}_{\alpha}dx^{\alpha}=-\frac{af_{,t}}{\sqrt{1-f_{,\chi}^{2}(1-\chi^{2})}}\,dt\,,
\end{equation}
which is well defined only if $1-f_{,\chi}^{2}(1-\chi^{2})>0$. Out
of this covector we can find its associate induced metric, $\mathfrak{h}_{\alpha\beta}=g_{\alpha\beta}+\mathfrak{n}_{\alpha}\mathfrak{n}_{\beta}$,
its extrinsic curvature tensor, $K_{\alpha\beta}=\mathfrak{h}_{\alpha}{}^{\rho}\mathfrak{h}_{\beta}{}^{\sigma}\nabla_{\rho}\mathfrak{n}_{\sigma}$
as well as its trace as $K=\mathfrak{h}^{\alpha\beta}K_{\alpha\beta}$.
Finally we can write
\begin{equation}
-K=\frac{(\chi^{2}-1)f_{,\chi\chi}}{a\,[1+f_{,\chi}^{2}\,(\chi^{2}-1)]^{\frac{3}{2}}}-\frac{3a_{,f}}{\sqrt{1+f_{,\chi}^{2}\,(\chi^{2}-1)}\,a^{2}}+\frac{2\left((\chi^{2}-1)^{2}f_{,\chi}^{2}+\frac{3\chi^{2}}{2}-1\right)f_{,\chi}}{[1+f_{,\chi}^{2}\,(\chi^{2}-1)]^{\frac{3}{2}}\chi\,a}=-K_{\infty}\,.\label{eq:K_K0_f}
\end{equation}
We will solve this differential equation
numerically later on, provided we also give the function $a=a(f)$ to be determined in the following. As for now, we will only assume there is such a solution. Once more, we need this equation to hold as we want to embed this GR solution in VCDM, and in this case, we need to require that $K=K_{\infty}$. 

We can now proceed by realizing that the hypersurface where the junction
conditions take place can be described from the inside-metric point
of view as happening for
\begin{equation}
\Phi(x^{\mu})=\chi-\chi_{s}=0\,,
\end{equation}
where a constant $\chi_{s}$ denotes the star surface, so that the unit
vector normal to this hypersurface can be written as
\begin{equation}
n_{\alpha}dx^{\alpha}=\frac{a}{\sqrt{1-\chi^{2}}}\,d\chi\,.
\end{equation}
Out of this covector one can define the projector $h_{\alpha}{}^{\beta}=\delta_{\alpha}{}^{\beta}-n_{\alpha}n^{\beta}$,
and the extrinsic curvature $\mathcal{K}_{\mu\nu}$. Also for the
inside metric we can define three vectors
\begin{equation}
e_{a}^{\mu}=\frac{\partial x^{\mu}}{\partial y^{a}}\,,\qquad{\rm where}\qquad y^{a}\in\{t,z,\theta_{2}\}\,,
\end{equation}
or
\begin{equation}
e_{t}^{\mu}\partial_{\mu}=\partial_{t}\,,\qquad e_{z}^{\mu}\partial_{\mu}=\partial_{z}\,,\qquad e_{\theta_{2}}^{\mu}\partial_{\mu}=\partial_{\theta_{2}}\,.
\end{equation}
By doing this, out of the twelve scalars
\begin{equation}
h_{ab}=h_{ba}=\left.h_{\mu\nu}e^{\mu}{}_{a}e^{\nu}{}_{b}\right|_{\chi=\chi_{s}}\,,\qquad\mathcal{K}_{ab}=\mathcal{K}_{ba}=\left.\mathcal{K}_{\mu\nu}e^{\mu}{}_{a}e^{\nu}{}_{b}\right|_{\chi=\chi_{s}}\,,
\end{equation}
we can prove that for the inside metric the only nonzero terms are
the following ones
\begin{eqnarray}
h_{tt} & = & -a\bigl(f(t,\chi_{s})\bigr){}^{2}f_{,t}(t,\chi_{s})^{2}\,,\\
h_{zz} & = & \frac{a\bigl(f(t,\chi_{s})\bigr){}^{2}\chi_{s}^{2}}{1-z^{2}}\,,\\
h_{\theta_{2}\theta_{2}} & = & a\bigl(f(t,\chi_{s})\bigr){}^{2}\chi_{s}^{2}\,(1-z^{2})\,,\\
\mathcal{K}_{zz} & = & \frac{a\bigl(f(t,\chi_{s})\bigr)\chi_{s}\sqrt{1-\chi_{s}^{2}}}{1-z^{2}}\,,\\
\mathcal{K}_{\theta_{2}\theta_{2}} & = & a\bigl(f(t,\chi_{s})\bigr)\chi_{s}\sqrt{1-\chi_{s}^{2}}(1-z^{2})\,.
\end{eqnarray}
It should be noticed that the component $\mathcal{K}_{tt}$ vanishes
for the inside metric.

\section{Matching conditions\label{sec:Matching-conditions}}

\subsection{Israel junction conditions}

We now impose the standard Israel junction conditions at the surface of the star. This standard treatment in GR is justified also in VCDM since (i) we have set the trace of the extrinsic curvature of the constant-$t$ hypersurface to the same constant value $K_{\infty}$ everywhere in both sides of the surface of the star; and (ii) we shall in the next subsection require the continuity of the unit normal to the constant-$t$ hypersurface across the surface of the star as (\ref{eqn:continuity-normal1})-(\ref{eqn:continuity-normal2}) below.

For the outside metric let us first consider the following two elements

\begin{eqnarray}
h_{zz}^{+} & = & \frac{R(t)^{2}}{1-z^{2}}\,,\\
h_{\theta_{2}\theta_{2}}^{+} & = & R(t)^{2}\,(1-z^{2})\,.
\end{eqnarray}
If we match them with the inside metric we find
\begin{equation}
R(t)=\chi_{s}\,a\bigl(f(t,\chi_{s})\bigr)\,.
\end{equation}

Let us now consider the expression
\begin{equation}
\left.\mathcal{K}_{zz}^{+}\right|_{r=R(t)}=\left.\mathcal{K}_{zz}^{-}\right|_{\chi=\chi_{s},a=R/\chi_{s}}\,.
\end{equation}
Then one finds a quadratic equation for $\dot{T}$ as
\begin{equation}
\dot{T}^{2}+\mathcal{A}(t)\,\dot{T}+\mathcal{B}(t)=0\,.\label{eq:Kzz_dT2}
\end{equation}
By taking its time derivative, we can also find an equation for $\ddot{T}$.
Eq.\ (\ref{eq:Kzz_dT2}) can be solved algebraically for $\dot{T}$
as 
\begin{equation}
\dot{T}=-\frac{\sqrt{1-\chi_{s}^{2}}\,\sqrt{1-\chi_{s}^{2}+\bar{\beta}^{2}-\bar{F}}\,\sqrt{\bar{F}}-(1-\chi_{s}^{2}+\bar{\beta}^{2}-\bar{F})\bar{\beta}}{(\bar{F}-\bar{\beta}^{2})(1-\chi_{s}^{2}+\bar{\beta}^{2}-\bar{F})\,\bar{g}_{2}}\,\dot{R}+\frac{\bar{g}_{1}\dot{\kappa}}{\bar{g}_{2}}\,,
\label{eqn:Tdot}
\end{equation}
where a bar indicates that the function is evaluated at $r=R(t)$,
e.g.\ $\bar{F}=F\bigl(t,r=R(t)\bigr)$. We have picked up the solution
which for vanishing $\beta$, leads to $\dot{T}>0$ for $\dot{R}<0$.
Here and in the following we always assume that $\kappa(t)>0$, so that
$\bar{\beta}>0$. On considering the possible value for $R(t)=R_{h}$
for which $\sqrt{\bar{F}}=\bar{\beta}>0$, then at this instant of time
the solution would be crossing the horizon. However, the quantity
$\dot{T}/\dot{R}$ would still remain finite as
\begin{equation}
\dot{T}-\frac{\bar{g}_{1}\dot{\kappa}}{\bar{g}_{2}}=-\frac{(1-\chi_{s}^{2}+\bar{\beta}^{2})}{(1-\chi_{s}^{2})\left[\sqrt{\bar{F}}\sqrt{1+\frac{\bar{\beta}^{2}-\bar{F}}{1-\chi_{s}^{2}}}\,+\left(1+\frac{\bar{\beta}^{2}-\bar{F}}{1-\chi_{s}^{2}}\right)\bar{\beta}\right]}\frac{\dot{R}}{\bar{g}_{2}}\,.
\end{equation}
On using the obtained relations for $\dot{T}$ and $\ddot{T}$, we
also find that
\begin{equation}
\mathcal{K}_{tt}^{+}=0\,,
\end{equation}
satisfying automatically the matching conditions.

At this level all the components of the extrinsic curvature tensor are equal
for both the inside/outside metrics. However, we still need to match
\begin{equation}
\mathcal{E}_{tt}\equiv h_{tt}^{+}-\left.h_{tt}^{-}\right|_{a=R/\chi_{s}}=0\,.
\end{equation}
Since $R(t)=\chi_{s}\,a[f(t,\chi_{s})]$, we can replace
\begin{equation}
\dot{R}=\frac{dR}{dt}=\frac{dR}{da}\,\frac{da}{df}\,f_{,t}=\chi_{s}\,a_{,f}\,f_{,t}\,.
\end{equation}
This relation, together with the one obtained before for $\dot{T}$
gives the following constraint
\begin{equation}
\mathcal{E}_{tt}=\frac{af_{,t}^{2}\left(a^{4}K_{\infty}^{2}\chi_{s}^{3}-9a^{2}\chi_{s}^{3}-9a_{,f}^{2}\chi_{s}^{3}+9ar_{H}\right)}{a^{3}K_{\infty}^{2}\chi_{s}^{3}-9a\chi_{s}^{3}+9r_{H}}=0\,.
\end{equation}
This constraint can be solved as
\begin{equation}
\frac{3a_{,f}^{2}}{a^{4}}+\frac{3}{a^{2}}=\frac{K_{\infty}^{2}}{3}+\frac{3r_{H}}{a^{3}\chi_{s}^{3}}\,.\label{eq:fried_int}
\end{equation}
This equation is the equation which determines $a$ as a function
of $f$, and it can be recognized as being the Friedmann equation
of a closed universe with a cosmological constant and a dust component
which can be written as
\begin{equation}
\frac{\rho_{{\rm dust}}}{\Mpl^{2}}=\frac{3r_{H}}{a^{3}\chi_{s}^{3}}=\frac{3(2GM)}{R^{3}}\,,\qquad{\rm or}\qquad M=\frac{4}{3}\pi\rho_{{\rm dust}}R^{3}\,.
\end{equation}
If we define $a_{\rm max}$ as the maximum value of $a$, in this case $a_{,f}(a=a_{\rm max})=0$, which also implies that $\dot{R}=0$. In this case, on the maximum, we also find
\begin{equation}
\frac{3r_{H}}{R_{\rm max}^{3}}=\frac{3r_{H}}{a_{\rm max}^{3}\chi_{s}^{3}}=\frac{3}{a_{\rm max}^{2}}-\frac{K_{\infty}^{2}}{3}=\frac{3\chi_{s}^{2}}{R_{\rm max}^{2}}-\frac{K_{\infty}^{2}}{3}\,,
\end{equation}
which sets the value of $\chi_{s}$ as
\begin{equation}
\chi_{s}^{2}=\frac{r_{H}}{R_{\rm max}}+\frac{K_{\infty}^{2}R_{\rm max}^{2}}{9}\,.
\end{equation}

\subsection{Absence of cusp of $\mathcal{T}$-constant surface}

As we can see from the VCDM action (\ref{vcdm-action2}) introduced in Section~\ref{sec:vcdm-action}, in VCDM we need to add an extra junction condition.  In fact, in VCDM there is a field $\mathcal{T}$ which is required to be timelike everywhere (and can be always fixed to be $\mathcal{T}=t$).  Hence $\mathcal{T}$ has a non-trivial profile and we need this field to be continuous at the
surface of the star. Out of $\mathcal{T}$, we define 
\begin{equation}
\mathcal{A}=\frac{1}{\sqrt{-\partial_{\rho}\mathcal{T}g^{\rho\sigma}\partial_{\sigma}\mathcal{T}}}\,,
\end{equation}
and the unit vector normal to constant-$\mathcal{T}$ hypersurfaces as 
\begin{equation}
\mathfrak{n}_{\mu}=-\mathcal{A}\,\partial_{\mu}\mathcal{T}\,,
\end{equation}
which on shell satisfies the following equation of motion
\begin{equation}
\nabla_{\mu}\mathfrak{n}^{\mu}=-\frac{3}{2}\,\lambda-\phi\,,
\end{equation}
where $\lambda$ and $\phi$ being the other two auxiliary field of
VCDM. For a solution of VCDM which is also a solution of GR both $\lambda$
and $\phi$ need to be constant. Then, let us consider both inside/outside
the solution with $\lambda$ and $\phi$ being constants everywhere.
Then, on choosing Gaussian normal coordinates about the matching hypersurface
we find
\begin{equation}
\partial_{l}[\sqrt{|h|}\mathcal{A}\partial^{l}\mathcal{T}]+\partial_{s}[\sqrt{|h|}\mathcal{A}h^{sr}\partial_{r}\mathcal{T}]=\left(\frac{3}{2}\,\lambda+\phi\right)\sqrt{|h|}\,,
\end{equation}
where $l$ is a coordinate orthogonal to the hypersurface. On integrating
about a thin layer in $l$ and since there are no singular terms we
find
\begin{equation}
\left.\sqrt{|h|}\mathcal{A}\partial^{l}\mathcal{T}\right|_{+}=\left.\sqrt{|h|}\mathcal{A}\partial^{l}\mathcal{T}\right|_{-},
\end{equation}
which leads to
\begin{equation}
\left.\mathcal{A}\partial^{l}\mathcal{T}\right|_{+}=\left.\mathcal{A}\partial^{l}\mathcal{T}\right|_{-},
\end{equation}
or, if $\mathcal{T}=t$, the previous relation can be rewritten as
\begin{equation}
[\mathcal{A}n^{\mu}\partial_{\mu}t]=\left[\frac{n^{\mu}\partial_{\mu}t}{\sqrt{-h^{00}-(n^{\sigma}\partial_{\sigma}t)^{2}}}\right]=0\,,\qquad{\rm or}\qquad[n^{\mu}\partial_{\mu}t]=0\,,
\end{equation}
having used the continuity of the induced metric $h_{\mu\nu}$. This
analysis shows that we need to impose one extra non-trivial matching
condition. In VCDM, we choose $\mathcal{T}=t$ (both inside and outside)
so that we have
\begin{align}
[n^{\mu}\mathfrak{n}_{\mu}] & =-\mathcal{A}[n^{\mu}\partial_{\mu}t]=0\,, \label{eqn:continuity-normal1}\\{} 
[e^{\mu}{}_{a}h_{\mu}{}^{\nu}\mathfrak{n}_{\nu}] & =-\mathcal{A}[e^{\mu}{}_{a}\partial_{\mu}t]=0\,.\label{eqn:continuity-normal2}
\end{align}
In summary, the normal to the $\mathcal{T}$-constant hypersurface is the same inside/outside the hypersurface. These conditions basically ensure the absence of cusp of the $\mathcal{T}$-constant hypersurface at the intersection with the surface of the star.

\section{Analytical and numerical insight on the collapse\label{sec:Analytical-and-numerical}}

In what follows, we will neglect the effect of the cosmological constant, because its value only gives tiny corrections to the numerical results and will not change the qualitative picture. This leads to considering $K_{\infty}=0$. Before focusing into the numerics we want to explicitly write down the condition  
\begin{equation}
[n^{\mu}\partial_{\mu}t]=0\,,
\end{equation}
in our case. Let us also redefine for later convenience
\begin{equation}
\kappa(t)=\kappa_{+}\kappa_{1}(t)\,,
\end{equation}
where we remind the reader that (neglecting the effective cosmological constant)
\begin{equation}
F=1-\frac{r_{H}}{r}+\frac{\kappa(t)^{2}}{r^{4}}\,,\qquad\kappa_{+}\equiv\frac{3}{16}\sqrt{3}r_{H}^{2}\,,
\end{equation}
so that $\kappa_{1}(t)$ becomes dimensionless. Then this last constraint
$[n^{\mu}\partial_{\mu}t]=0$ can be used as to constrain the dynamics
of $\kappa_{1}(t)$. The acceptable solutions can be written as
\begin{equation}
\kappa_{1}=\pm\frac{4\sqrt{3}\{\chi_{s}[1-\cos f(t,\chi_{s})]+f_{,\chi}(t,\chi_{s})\,(\chi_{s}^{2}-1)\sin f(t,\chi_{s})\}[1+\cos f(t,\chi_{s})]{}^{\frac{3}{2}}}{9\chi_{s}^{4}\sqrt{1-\cos f(t,\chi_{s})}\,\sqrt{1+(\chi_{s}^{2}-1)f_{,\chi}(t,\chi_{s})^{2}}}\,.\label{eq:kappa1}
\end{equation}
This shows that once the solution $f(t,\chi)$ is known, i.e., if we obtain 
an appropriate slicing which satisfies all necessary conditions, the behavior
of $\kappa_{1}$ is also known.

For the internal solution, in the absence of a cosmological constant,
we have already found in Eq.\ (\ref{eq:fried_int}) that
\begin{equation}
\frac{a_{,f}^{2}}{a^{4}}+\frac{1}{a^{2}}=\frac{r_{H}}{a^{3}\chi_{s}^{3}}\,,
\end{equation}
which is solved by 
\begin{equation}
a=\frac{1}{2}\,a_{\rm max}\,[1+\cos f]\,,\qquad0\le f<\pi\,,\qquad{\rm and}\qquad a_{\rm max}=\frac{r_{H}}{\chi_{s}^{3}}\,,
\end{equation}
having fixed the initial condition such that $a(f=0)=a_{\rm max}$. The
singularity of $a=0$ is reached whenever $f=\pi$. Using this solution,
we want to discuss, given a value of $\chi_{s}$, the values of $f$ for
which the solution hits the apparent horizon. In fact, the apparent
horizon is located at the solution for $R(t)=r_{H}$. In other words,
we want to find the value of $f=f_{H}$ at which this happens. Then we
need to solve
\begin{equation}
\frac{r_{H}}{\chi_{s}}=a(f=f_{H})\,,
\end{equation}
which then gives the following result
\begin{equation}
f=f_{H}\equiv\arccos(2\chi_{s}^{2}-1)\,.
\end{equation}
Thus, for a given value of $\chi_{s}$, the part of the interior solution with $f(t,\chi_{s})>f_{H}(\chi_{s})$ is inside the apparent horizon, as seen from the outer space.

Next, we substitute the function $a(f)$ inside Eq.\ (\ref{eq:K_K0_f})
as to find an elliptic equation $f$ which can be rewritten as
\begin{equation}
f_{,\chi\chi}=\frac{2(1-\chi^{2})\,f_{,\chi}^{3}}{\chi}-\frac{3f_{,\chi}^{2}\sin f}{1+\cos f}+\frac{(2-3\chi^{2})f_{,\chi}}{\chi\,(\chi^{2}-1)}-\frac{3\sin f}{(\chi^{2}-1)\,(1+\cos f)}\,.\label{eq:fcc}
\end{equation}
This equation does not involve any time derivative and thus can be solved on each constant-$t$ hypersurface by imposing the following boundary conditions
\begin{equation}
f(t,\chi=0)=f_{0}(t)\,,\qquad f_{,\chi}(t,\chi=0)=0\,,\label{eq:fcc-bc}
\end{equation}
where here the suffix 0 stands for the center of the dust cloud. The second condition is imposed as to make the slicing non singular at the center of the star.

In case the collapse runs with $\chi\leq\chi_{s}\ll1$, i.e.\ for
$r_{H}\ll R_{\rm max}$, the Taylor series in $\chi$ provides an excellent approximation for the
solution of $f$ as
\begin{equation}
f=f_{0}(t)+\frac{1}{2}\left(\frac{\sin f_{0}(t)}{1+\cos f_{0}(t)}\right)\chi^{2}+\dots+\mathcal{O}[\chi^{14}]\,,\label{eq:approx_f}
\end{equation}
where we have set boundary conditions as to have $\lim_{\chi\to0}f_{,\chi}=0$.
We can also see that $\lim_{f_{0}\to0}f=0$. In the remaining section,
in order to have a firm analytical insight, we will use this
approximate solution\footnote{Numerically, the Taylor series in this solution converges slowly because of the negative powers of $1+\cos f_{0}$. This solution nonetheless gives the same qualitative results as the numerical results, for values of $0<\chi_{s}\lesssim10^{-2}$.}. On specifying the function $f_{0}(t)\equiv f(t,0)$, we will also
solve Eq.\ (\ref{eq:fcc}) numerically (for values of $\chi_{s}$
closer to unity). In any case the dependence of $f$ on time is only
through the function $f_{0}(t)$.

Then the function $f(t,\chi)$, both analytically and numerically, does depend on $t$ only through $f_{0}(t)$, therefore we have
\begin{equation}
f(t,\chi)=\tilde{f}(f_{0}(t),\chi)\,. \label{eqn:f-f0andchi}
\end{equation}
This is because, as already mentioned, (\ref{eq:fcc}) with (\ref{eq:fcc-bc}) can be integrated on each constant-$t$ hypersurface. Hereafter, for simplicity of notation we express the function on the right hand side of (\ref{eqn:f-f0andchi}) as $f((f_{0}(t),\chi))$ by omitting the tilde. 

In the remaining part we will only consider the positive solution
for $\kappa_{1}$ defined inside Eq.\ (\ref{eq:kappa1}), so that
we have
\begin{equation}
\kappa_{1}=\tilde{\kappa}_{1}(f_{0}(t),\chi_{s})\,.
\end{equation}
Hereafter, for simplicity we express the function on the right hand side as $\kappa_{1}((f_{0}(t),\chi))$ by omitting the tilde. In this section, when we do numerics, we will set $f_{0}(t)=t$, as to try to reach any point of the solution (making omitting the tilde justified). We know that in any case $0\leq f<\pi$. However, in the analytic description that follows, we will study the behavior of the solution without specifying any explicit dependence of $f_{0}$ on time. 

\subsection{Vanishing lapse as final point of the collapse}

\begin{figure}
\includegraphics[width=9cm]{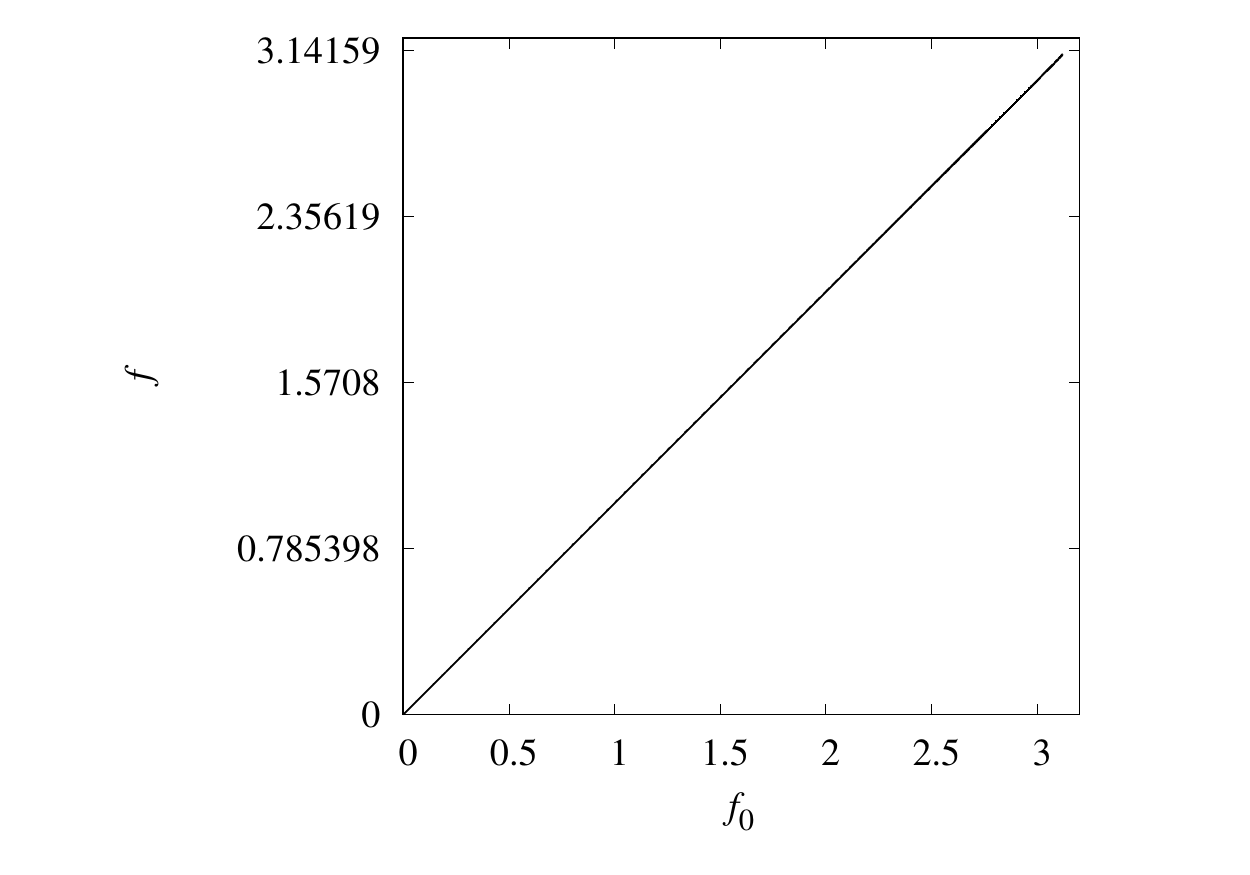}\!\!\!\!
\hspace{-1cm}
\includegraphics[width=9cm]{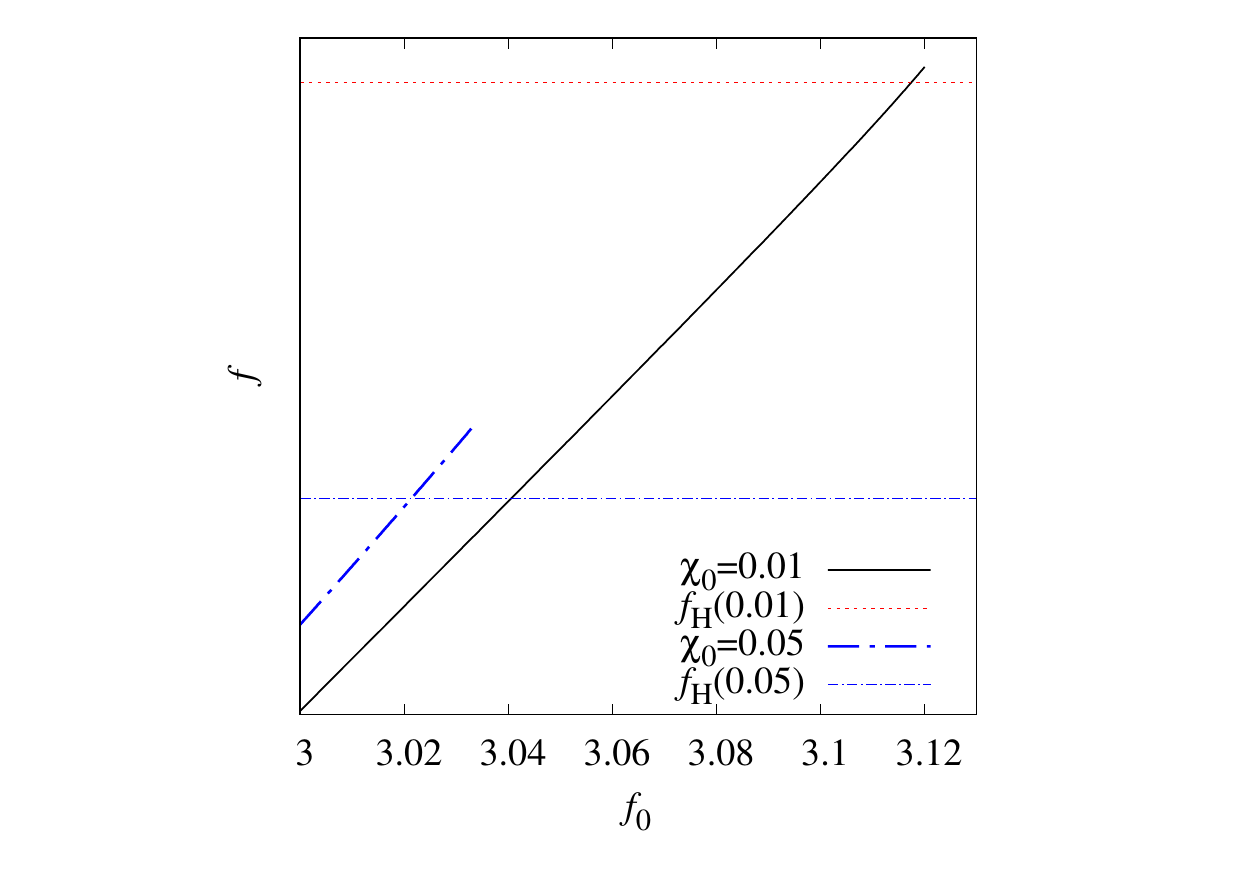}
\caption{Plot of the function $f_{s}=f(f_{0}(t),\chi_{s})$. In the
  left panel, we show that in the allowed range of $f_{0}$, $f_{s}$
  and ${f_{s}}_{,f_{0}}$ are always finite for a fixed value of
  $\chi_{s}=0.01$.  This behavior holds for other values of $\chi_{s}$
  as well. In the right figure, we show that the collapse ends inside
  the apparent horizon of the metric, this because $f>f_{H}$. Once
  more this behavior hold for all values of $\chi_{s}$ such that
  $0<\chi_{s}<1$.\label{fig:Plot-of-f}}
\end{figure}

Let us consider once more the lapse for the exterior solution (for
any value of $r$). Then one finds
\begin{equation}
N(t,r)=\alpha(t,r)\,\dot{T}(t)\,,
\end{equation}
where
\begin{eqnarray}
\alpha & = & \left(1-\frac{\dot{\kappa}}{\dot{T}}\int_{\infty}^{r}\frac{1}{r_{1}^{2}}F(t,r_{1})^{-3/2} \mathrm{d}r_{1}\right)\sqrt{F}\,,\label{eq:lapse_fin}\\
F & = & 1-\frac{r_{H}}{r}+\frac{\kappa(t)^{2}}{r^{4}}\,,
\end{eqnarray}
so that $\lim_{r\to\infty}\alpha=1$. We also set boundary conditions
for $N(t,r)$ so that $\lim_{r\to\infty}N(t,r)=1$. This means that
$t$ is the cosmological time, that is the time as measured by an
observer far away from the black hole. In this case we have that 
\begin{equation}
\dot{T}=1\,.
\end{equation}
This, combined with (\ref{eqn:Tdot}), leads to
\begin{align}
1 & =-\frac{[\kappa^{2}R_{\rm max}+R^{4}(R_{\rm max}-r_{H})]RR_{,f_{0}}\dot{f}_{0}}{\sqrt{R^{4}-R^{3}r_{H}+\kappa^{2}}\,[\sqrt{R^{4}-R^{3}r_{H}+\kappa^{2}}\,\sqrt{R}\,\sqrt{R_{\rm max}-r_{H}}\,\sqrt{r_{H}(R_{\rm max}-R)}+r_{H}\kappa\,(R_{\rm max}-R)]}\nonumber \\
 & +\left({\displaystyle \int_{\infty}^{R}}
 \frac{1}{r_{1}^{2}}F(t,r_{1})^{-3/2} 
 {d}r_{1}\right)\kappa_{,f_{0}}\dot{f}_{0}\,.
\end{align}
On doing this, $f_{0}(t)$ depends on the collapse parameters: indeed
this last condition corresponds to an ODE for $f_{0}$, which can
be written as
\begin{equation}
\dot{f}_{0}=-\frac{1}{\left({\displaystyle {\int}_{0}^{1/R}}
F\left(t, {1\over u}\right)^{-3/2}
{d}u\right)\kappa_{,f_{0}}+\frac{[\kappa^{2}R_{\rm max}+R^{4}(R_{\rm max}-r_{H})]RR_{,f_{0}}}{\sqrt{R^{4}-R^{3}r_{H}+\kappa^{2}}\,[\sqrt{R^{4}-R^{3}r_{H}+\kappa^{2}}\,\sqrt{R}\,\sqrt{R_{\rm max}-r_{H}}\,\sqrt{r_{H}(R_{\rm max}-R)}+r_{H}\kappa\,(R_{\rm max}-R)]}}\,,\label{eq:dot_f0}
\end{equation}
where we have changed the integration variable from $r_{1}$ to $u=1/r_{1}$.  We can insert this value of $\dot{f}_{0}$ into the expression in Eq.\ (\ref{eq:lapse_fin}) for the lapse $N$ (or $\alpha$) through $\dot{\kappa}=\dot{f}_0\kappa_{,f_{0}}$, where on the right hand side we have considered $\kappa$ as a function of $f_0$ and $\chi$, finding
\begin{eqnarray}
N&=&
\left[
1-\frac{\kappa_{,f_{0}}\displaystyle{ {\int}_{0}^{1/r}}
F\left(t, {1\over u}\right)^{-3/2} du}{\kappa_{,f_{0}}\displaystyle{ {\int}_{0}^{1/R}}
F\left(t, {1\over u}\right)^{-3/2} du+\frac{[{\kappa^{2}\over R^4}
R_{\rm max}+(R_{\rm max}-r_{H})]RR_{,f_{0}}}{\sqrt{F(t, R)}\,\left[\sqrt{F(t,R)} \,\sqrt{R}\,\sqrt{R_{\rm max}-r_{H}}\,\sqrt{r_{H}(R_{\rm max}-R)}+r_{H}{\kappa\over R^2}\,(R_{\rm max}-R)\right]}}
\right]
\nonumber 
\\
&&
\times \sqrt{F(\kappa,r)}\,.
\end{eqnarray}
The function $f_{,f_{0}}$ both analytically and numerically keeps
finite during the whole evolution, as it is shown in Fig.\ \ref{fig:Plot-of-f}.

Therefore for $\kappa_{1}^{2}>1$, that is for $\kappa>\kappa_{+}$
(assuming positive $\kappa_{1}$), the integral is always well behaved.
If $0<\kappa_{1}<1$ the integral and the squared roots are well
behaved as long as $R(t)$ does not coincide with one of the two real
roots of $F(t,r)=1-r_{H}/r+\kappa(t)^{2}/r^{4}$. In fact, also the
other term in the denominator, the one proportional to $R_{,f_{0}}$ tends
to blow up once more if $R(t)\neq0$ (or $f<\pi$) and $R(t)$ is such
that $F(t,r=R(t))=0$. Let us then consider the dynamics of
$\kappa_{1}$, where 
\begin{equation}
\kappa=\kappa_{+}\kappa_{1}\bigl(f(f_{0},\chi_{s})\bigr)\,,\qquad{\rm and}\qquad\kappa_{+}=\frac{3\sqrt{3}}{16}\,r_{H}^{2}\,.
\end{equation}
We find that it depends on the value of $\chi_{s}$. So let us now
study this behavior in some detail.

\subsubsection{Small values of $\chi_{s}$}

\begin{figure}[ht]
\includegraphics[width=10cm]{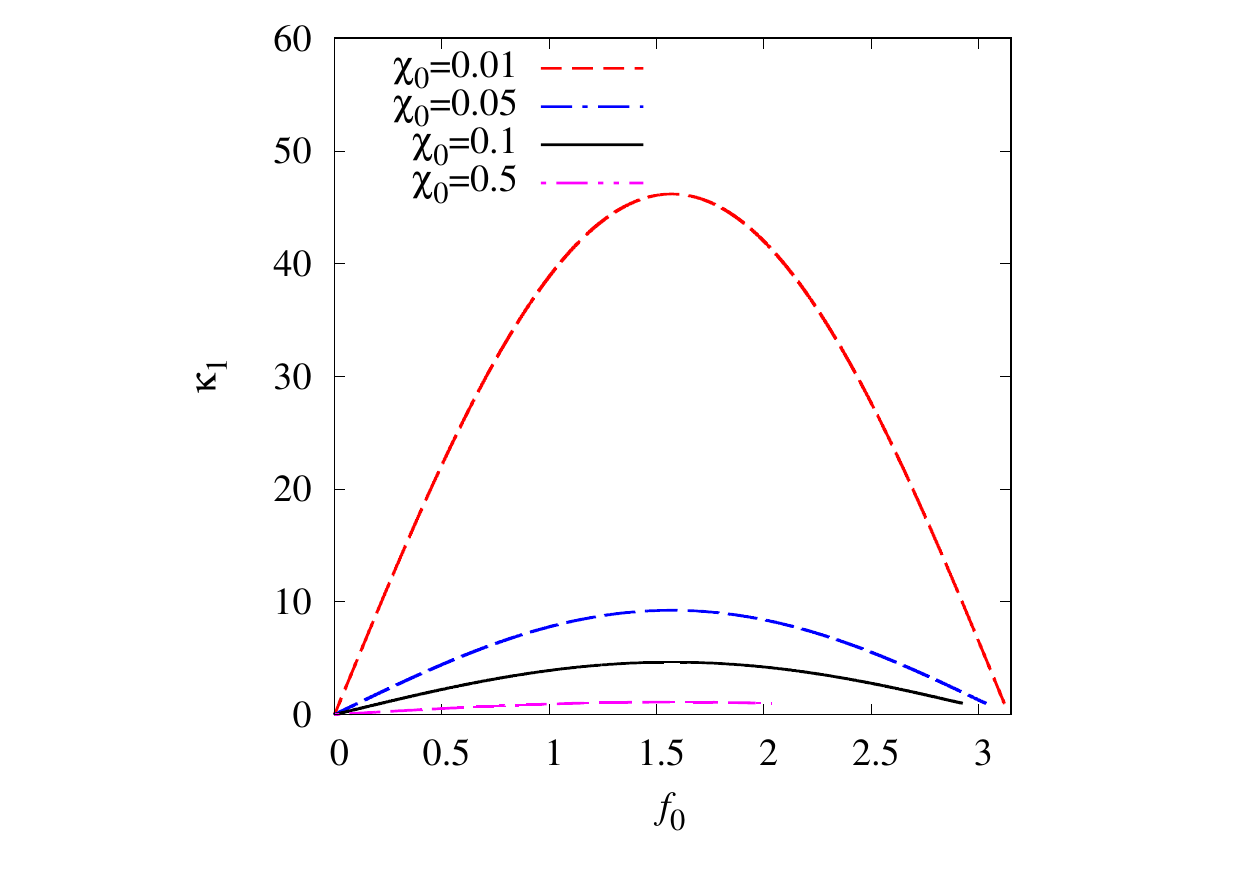}\!\!\!\!\hspace{-2cm}
\includegraphics[width=10cm]{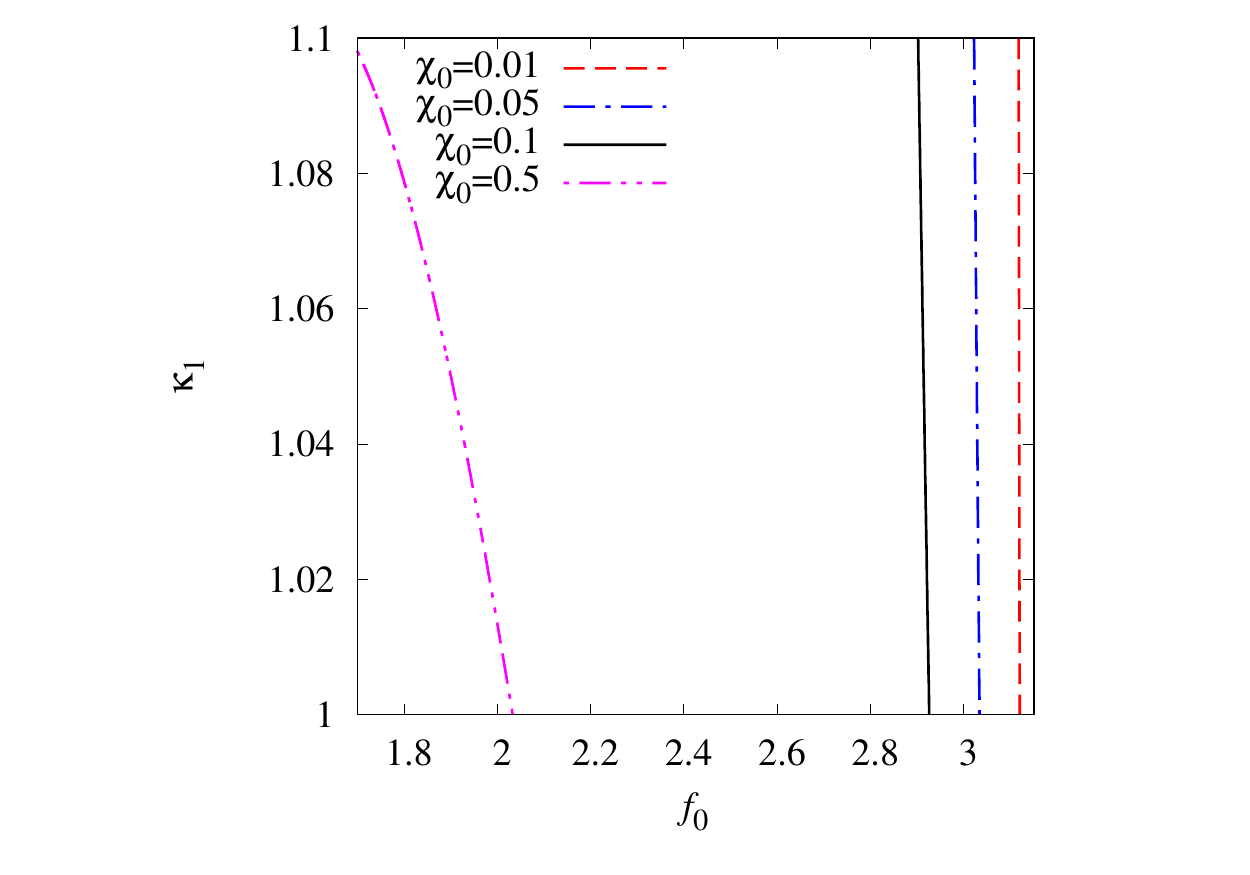}
\caption{Plots of $\kappa_{1}$ as a function of $f_{0}$ for different values
of $\chi_{s}$. In the left panel we can see that $\kappa_{1}$ starts
from zero, grows up to a maximum (which is larger than unity)
and then $\kappa_{1}$ decreases up to the point it reaches the value
of unity (as it can be seen in the right figure). This final value
of $\kappa_{1}$ being unity takes place when the solution has already
entered its own apparent horizon, as discussed previously 
 in Fig.\ \ref{fig:Plot-of-f}.
In other words, at horizon crossing, we find that $\kappa_{1}(t)>1$.
\label{fig:Plots-of-kappa1}}
\end{figure}

For some values of $\chi_{s}$, with $\chi_{s}\lesssim0.65$, we can
distinguish the following stages, which can be seen in Fig.\ \ref{fig:Plots-of-kappa1}.

\begin{figure}
\includegraphics[width=9cm]{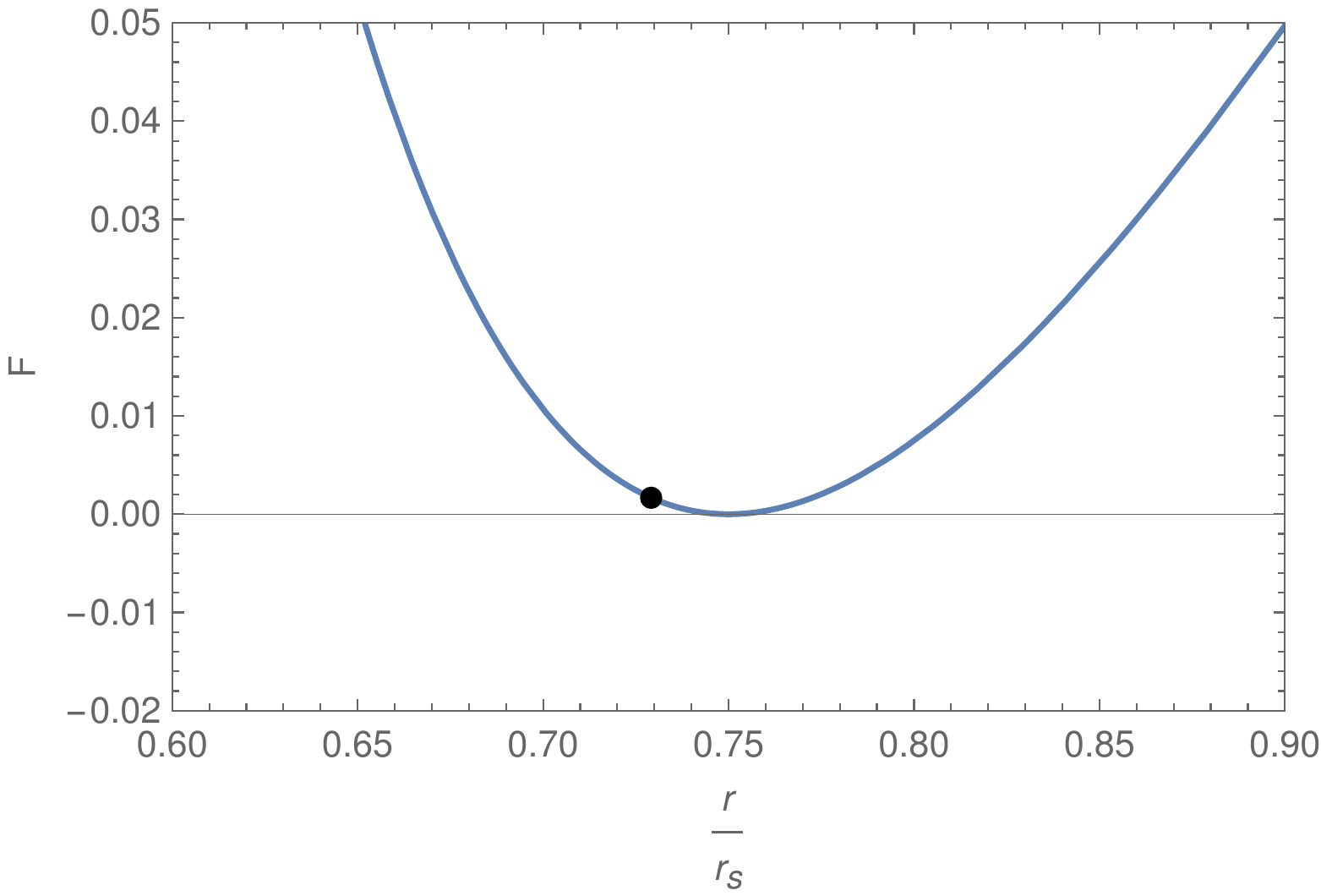}
\caption{Reaching the point at which $\dot{f}_{0}$ vanishes, for small values
of $\chi_{s}$. In fact, $\dot{f}_{0}\to0$ at a value of $f_{0}=f_{0F}$
such that $\kappa_{1}(f(f_{0F},\chi_{s}))=1$. In this case the integral
$\textcolor{gray}{\int}_{0}^{1/R(t)}F\left(t,{1\over u} \right)^{-3/2}
\textcolor{gray}{d}u$
diverges as the variable $u$ crosses the value $4/(3r_{H})$, since
$\kappa_{+}\to1$, and $R(f_{0F})<3r_{H}/4$ as shown as a dot in the plot.
\label{fig:Reaching-the-point-kappa1}}
\end{figure}

\begin{enumerate}
\item First stage: initial collapse. At this stage $\kappa_{1}(f_{0}=0)=0$
and $\kappa_{1,f}>0$. This stage does not show any problem as $R\approx R_{\rm max}$
and $F>0$. Soon $\kappa_{1}$ reaches values larger than unity (for
which $F$ cannot vanish any more), but $\kappa_{1,{f_0}}$ tends to reduce
up to a time (or a value of $f_{0}$) where it flips sign.
\item Second stage: $\kappa_{1,f_{0}}<0$ up to horizon crossing. After
$\kappa_{1}$ has reached a maximum (larger than unity) indeed $\kappa_{1}$
starts decreasing while remaining larger than unity. In this case
$F$ never vanishes. At horizon crossing, that is when $f=f_{H}$,
still $\kappa_{1}>1$, so the collapsed star enters its own horizon.
\item Third stage: soon after the solution enters its own horizon, then
$\kappa_{1}(f_{0})$ keeps decreasing up to the value of $f_{0}=f_{0F}$
at which $\kappa_{1}(f_{0}=f_{0F})=1$. In this case, $F$ vanishes
for the value of $r_{0F}$, $r_{0F}=3r_{H}/4$, such that $R(f_{0F})<r_{0F}<r_{H}$.
Indeed one can study both $R$ and $\kappa_{1}$ as functions of $f_{0}$.
One sees that $\kappa_{1}(f_{0})$ reduces but $F(f_{0},r=R(f_{0}))$
reaches its minimum when $\kappa_{1}(f_{0})$ is still larger than
unity. Then $R(f_{0})$ will be such that $F(f_{0},r=R(f_{0}))$ will
be located to the right of the the minimum of the function $F(f_{0},r)$.
However, as already stated above, $f_{0}$ reaches a point such that
$\kappa_{1}(f_{0F})=1$. At this time there exists an $r=r_{0F}$
at which $F(f_{0},r=r_{0F})=0$. This point makes the integral in
the denominator of $\dot{f}_{0}$ blow up and also makes $\dot{f}_{0}$
vanish. When this happens for some value of $f_{0}=f_{0F}$, one finds
that $\kappa_{1}=1$, however the surface of the star (represented as a dot in 
Fig.\ \ref{fig:Reaching-the-point-kappa1}) is located at $R(f_{0F})/r_{H}<3/4$.
\end{enumerate}
In Fig.\ \ref{fig:Reaching-the-point-kappa1} we plot the function
$F(t,r)$ (which depends on time via the function $\kappa(t)$) at
the instant of time when $\kappa_{1}$ becomes equal to unity as to
make the integral in Eq.\ (\ref{eq:dot_f0}) diverge. As a consequence,
as we will discuss below, as $\kappa_{1}\to1$ the lapse tends to
vanish, i.e.\ $N\to0$, and for the cosmological observer the singularities
(e.g.\ when $a$ vanishes) are never reached, as the solutions will
take an infinite time to reach $f_{0F}$, i.e.\ the left zero-$F$
point, as $\dot{f}_{0}\to0$.

\subsubsection{Large values of $\chi_{s}$}

\begin{figure}
\includegraphics[width=9cm]{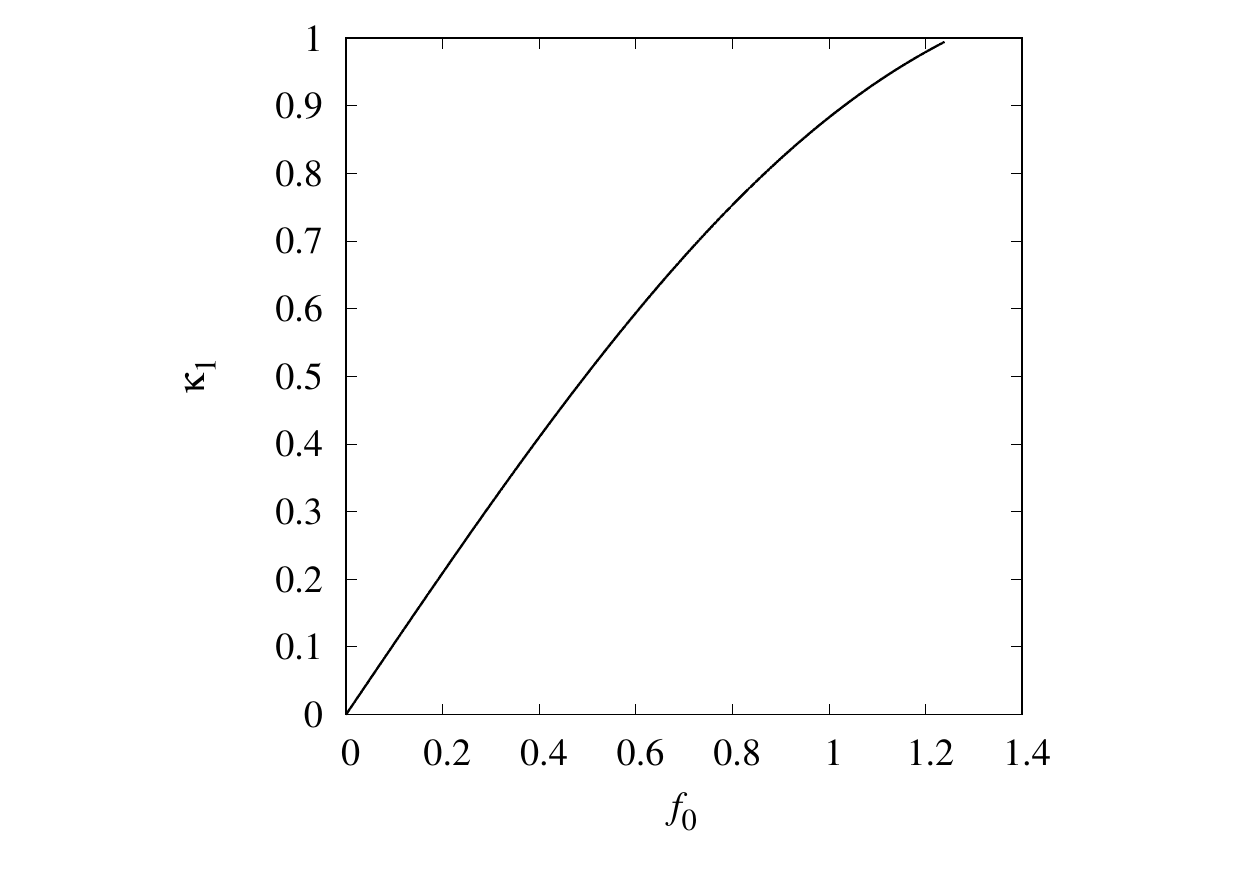}\!\!\!
\hspace{-2cm}
\includegraphics[width=9cm]{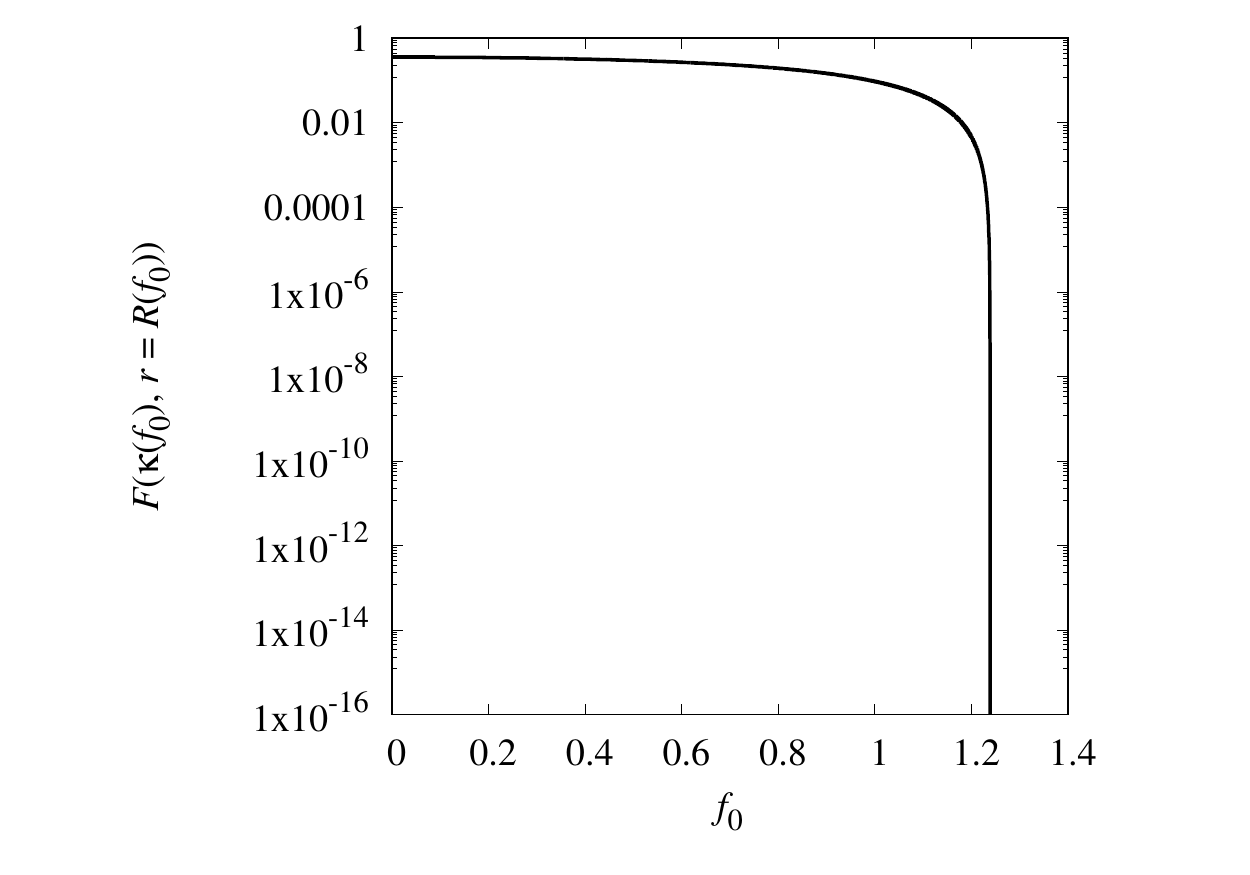}
\caption{In the left panel, we show a plot of the function $\kappa_{1}(f_{0})$
for a value of $\chi_{s}=0.8$. It is clear that although $\kappa_{1}$
increases, in the allowed domain of $f_{0}$ it never reaches unity.
In the right panel, we instead show, for the same value of $\chi_{s}$,
the behavior of $F(f_{0})$, and it is clear that it vanishes while
$0<\kappa_{1}<1$. At this point, that we name as $f_{0}=f_{0,+}$,
which still happens when the solution has already entered its apparent
horizon, we have that $\lim_{f_{0}\to f_{0,+}}F(\kappa_{1}(f_{0}),R(f_{0}))=0$,
and as such we have that $\lim_{f_{0}\to f_{0,+}}\dot{f}_{0}=0$.
\label{fig:Plot-of-f-large_chi}}
\end{figure}

For larger values of $\chi_{s}$, i.e.\ $0.65\apprle\chi_{s}<1$,
the dynamics of $\kappa_{1}(f_{0})$ does change and, in this case,
the collapse works in a different way. In fact, we have that $\kappa_{1}(f_{0})$
remains always smaller than unity. This seems to be giving problems
since the beginning of the collapse. However, the collapse starts
at $R(f_{0}=0)=R_{\rm max}>r_{H}$ so that at least initially $R_{\rm max}$ is
not one of the two roots of $F$. In other words, if $\kappa_{1}<1$,
$F$ can vanish but it does not as long as $R$ does not reach a minimum
critical value, $r_{1,+}=R(f_{0,+})$, at which $F=0$. We know that
if $0<\kappa_{1}<1$, the function $F$ vanishes at two values of
$r$, $r_{F\pm}$ so that $r_{F-}<\frac{3}{4}\,r_{H}<r_{F+}$ and
$r_{F+}-r_{F-}\to0$ as $\kappa_{1}\to1^{-}$. The collapse stops
when, at a given value of $f_{0}=f_{0,+}$, we have that $\kappa_{1}(f_{0,+})<1$
and $F(\kappa_{1}(f_{0,+}),R(f_{0,+}))=0$. In fact, we do have that,
because of the collapse, $R_{,f_{0}}<0$ and $R(f_{0})$ reaches the
larger root of $F$, $R=R(f_{0,+})$, before $\kappa_{1}$ reaches
unity, and this happens indeed at $f_{0}=f_{0,+}$. At this point
indeed once more both terms in the denominator of $\dot{f}_{0}$ in
Eq.\ (\ref{eq:dot_f0}) tend to diverge as $R$ approaches the largest
root of $F(\kappa_{1}(f_{0,+}),R(f_{0,+}))=0$. Fig.\ \ref{fig:Plot-of-f-large_chi} is devoted to
this case.

\subsection{Limiting surface}

Here we find the limiting surface, i.e.\ the surface with the largest
radius at which $N$ vanishes. We consider both the small and large
values for $\chi_{s}$.

\subsubsection{Small values of $\chi_{s}$}

Let us consider here the case of small $\chi_{s}$, and define
$f_{0F}$, that is the value of $f_{0}$ at which
$\kappa_{1}(f_{0F})=1$. Then let us also define
$R_{0F}=R(f_{0}=f_{0F}$). In general the lapse $N$ (or $\alpha$) is a
function of time (or $f_{0}$) and $r$ (as seen from the outer metric
point of view). Hence, let us evaluate the lapse at $f_{0}=f_{0F}$ but
for $r=r_{+}>R_{0F}$ at which $F(f_{0F},r_{+})=0$, that is
$N(f_{0F},r_{+})$ and $r_{+}=\frac{3}{4}\,r_{H}$. Notice that in this
case $F(f_{0F},r=R_{0F})\neq0$ in general. Then setting $F_+(t,r)
=F(t,r:\kappa=\kappa_+)$, we have
\begin{align}
& N(f_{0F},r_{+})  =\left(1-\kappa_{,f_{0}}\dot{f}_{0}\int_{\infty}^{r_{+}}\frac{1}{r_{1}^{2}}F(t,r_{1})^{-3/2}\mathrm{d}r_{1}\right)\sqrt{F(f_{0F},r_{+})}=\sqrt{F(f_{0F},r_{+})}\nonumber \\
 & \times\left(1-\frac{\kappa_{,f_{0}}\displaystyle{ {\int}_{0}^{1/r_+}}
 F_+\left(t, {1\over u}\right)^{-3/2} du }
 { \kappa_{,f_{0}} \displaystyle{ {\int}_{0}^{1/R}}F_+\left(t, {1\over u}\right)^{-3/2} du 
+\frac{[{\kappa_{+}^{2}\over R^4}+(R_{\rm max}-r_{H})]RR_{,f_{0}}}{\sqrt{F_+(t,R)}
\,[\sqrt{F_+(t,R)}\,\sqrt{R}
\sqrt{R_{\rm max}-r_{H}}\,\sqrt{r_{H}(R_{\rm max}-R)}+r_{H}{\kappa_{+}\over R^2}
\,(R_{\rm max}-R)]}}\right)
\nonumber \\
 & \approx\left(1-\frac{\displaystyle{ {\int}_{0}^{1/r_+}}F_+\left(t, {1\over u}\right)^{-3/2} du }{\left(\displaystyle{ {\int}_{0}^{1/r_+}}F_+\left(t, {1\over u}\right)^{-3/2} du+\displaystyle{ {\int}_{1/r_+}^{1/R_{0F}}}F_+\left(t, {1\over u}\right)^{-3/2} du\right)}\right)\sqrt{F_+(f_{0F},r_{+})}\nonumber \\
 & =\left(1-\frac{1}{1+C_{0}}\right)\sqrt{F_+(f_{0F},r_{+})}=\frac{\sqrt{F_+(f_{0F},r_{+})}}{2}=0\,,
\end{align}
because $F_+(r_{+})=0$, and 
\begin{equation}
C_{0}=\lim_{\epsilon\to0^{+}}\frac{\displaystyle{ {\int}_{1/(r_+-\epsilon)}^{1/R_{0F}}F_+\left(t, {1\over u}\right)^{-3/2}}\textcolor{gray}{d}u}{\displaystyle{ {\int}_{0}^{1/(r_++\epsilon)}}F_+\left(t, {1\over u}\right)^{-3/2} du}=1\,.
\end{equation}
This shows that for small $\chi_{s}$ we have $N(f_{0F},r_{+})=0$.

Let us now evaluate the lapse $N$ at $r=R_{0F}$, that is $N(f_{0F},R_{0F})$.
As a difference from the previous calculation, we have that $F(f_{0},R_{0F})\neq0$
but finite as seen also in Fig.\ \ref{fig:Reaching-the-point-kappa1}.
We have that
\begin{eqnarray}
N(f_{0F},R_{0F}) &
 =&\left(
 1-\kappa_{,f_{0}}\dot{f}_{0}\int_{\infty}^{R_{0F}}\frac{1}{r_{1}^{2}}
 F_+(t,r_{1})^{-3/2} \mathrm{d}r_{1}
 \right)\sqrt{F_+}\,,
 \nonumber \\
 &&
 \approx
 \left(
 1+\kappa_{,f_{0}}
 {1\over 
  \left(-\displaystyle{ \int_{0}^{1/R_{0F}}}
  F_+\left(t,{1\over u}\right)^{-3/2} du  
\right)
 \kappa_{,f_{0}}
 }
 \times
  \displaystyle{ \int_{0}^{1/R_{0F}}}
  F_+\left(t,{1\over u}\right)^{-3/2} du 
 \right)\sqrt{F_+}
  \nonumber \\
 && =0\,.
\end{eqnarray}
This shows that $N(f_{0F},R_{0F})=0$, with $R_{0F}<r_{+}$.

Finally, on evaluating the lapse $N$ at an intermediate point $R_{0F}<r_{0,+}<r_{+}$,
we find
\begin{align}
N(f_{0F},r_{0,+}) & =\left(1-\kappa_{,f_{0}}\dot{f}_{0}\int_{\infty}^{r_{0,+}}\frac{1}{r_{1}^{2}}F_+(t,r_{1})^{-3/2}\mathrm{d}r_{1}\right)\sqrt{F_+}\,,\nonumber \\
 & \approx\sqrt{F_+}\left[1-\frac{1}{\left(\displaystyle{ {\int}_{0}^{1/R_{0F}}}F_+\left(t,{1\over u}\right)^{-3/2} du \right)}
 \left(\displaystyle{ {\int}_{0}^{1/R_{0F}}}F_+\left(t,{1\over u}\right)^{-3/2} du 
 +\displaystyle{ {\int}_{1/R_{0F}}^{1/r_{0,+}}}F_+\left(t,{1\over u}\right)^{-3/2} du \right)\right]\nonumber \\
 & \approx\left(1-1\right)\sqrt{F(r_{0,+})}=0\,,
\end{align}
since
\begin{equation}
\frac{\displaystyle{ {\int}_{1/R_{0F}}^{1/r_{0,+}}}F_+\left(t,{1\over u}\right)^{-3/2} du}{\displaystyle{ {\int}_{0}^{1/R_{0F}}}F_+\left(t,{1\over u}\right)^{-3/2} du }=0\,,
\end{equation}
as the numerator is finite. Then this means that, at $f_{0}=f_{0F}$,
$N\to0$ from the origin up to $r_{+}=\frac{3}{4}\,r_{H}$. So the
limiting surface is $r=r_{+}$. This result then leads to the conclusion
that the metric, at the end of the collapse (which takes an infinite
time $t$) coincides with the spherical symmetric static solution
having $\kappa=\kappa_{+}$.

\begin{figure}
\includegraphics[width=11cm]{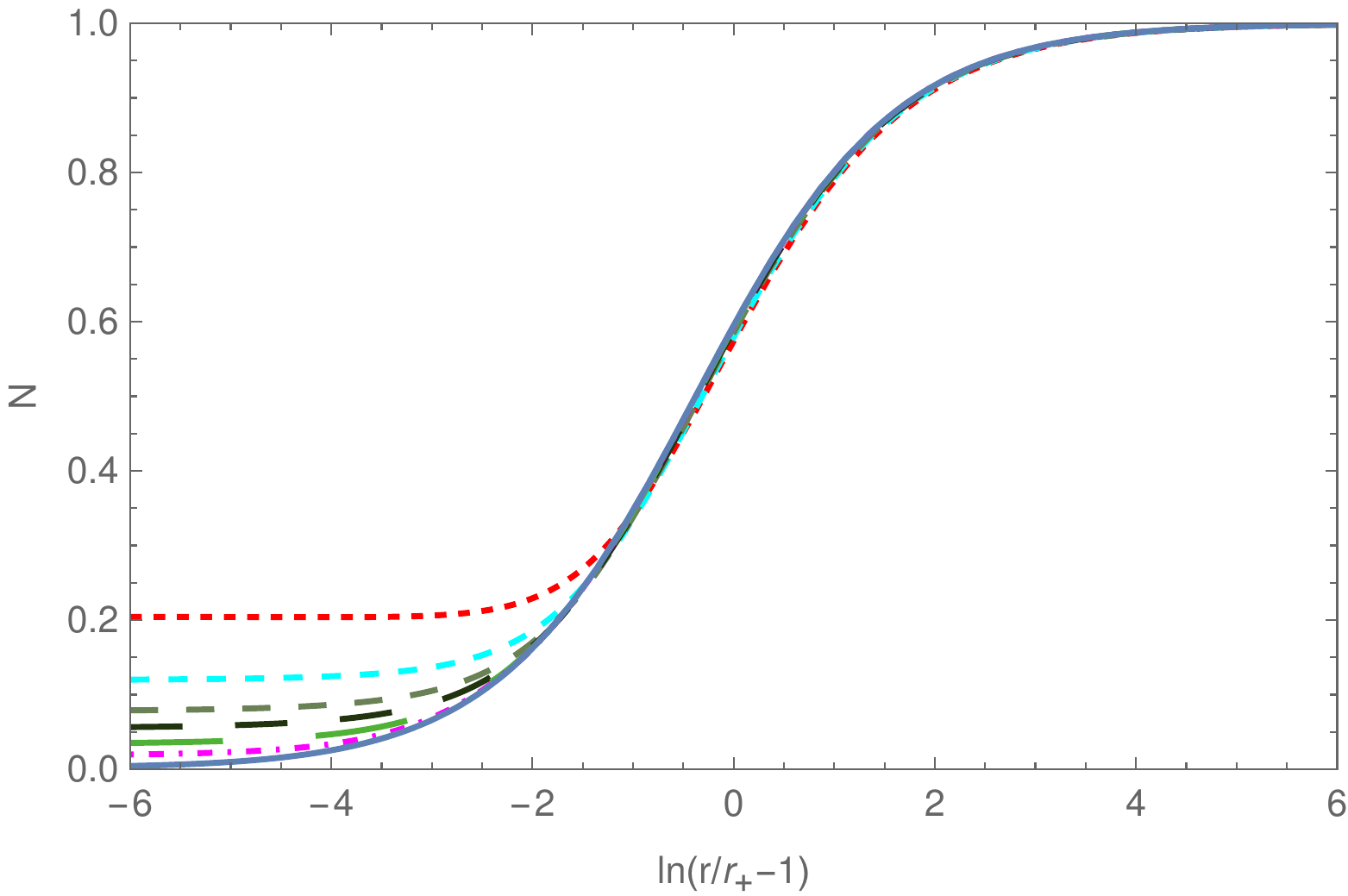}
\caption{
The lapse function $N$ in  the outer region ($r>r_{+}={3\over 4}r_{H}$) 
 is plotted  for several values of $f_0$ ($f_0=3.116$, 3.118, 3.119, 3.1195, 3.11986, 3.12001, 3.12006292)
 as $f_0\rightarrow f_{0F}$.}
\label{fig:The-lapse-N-small_chi}
\end{figure}

In particular, on plotting the $N$ as a function of $r$ for the
outer metric we can see that $N\to0$ as $r\to r_{+}$ as shown in
Fig.\ \ref{fig:The-lapse-N-small_chi}.

\subsubsection{Large values of $\chi_{s}$}

\begin{figure}
\includegraphics{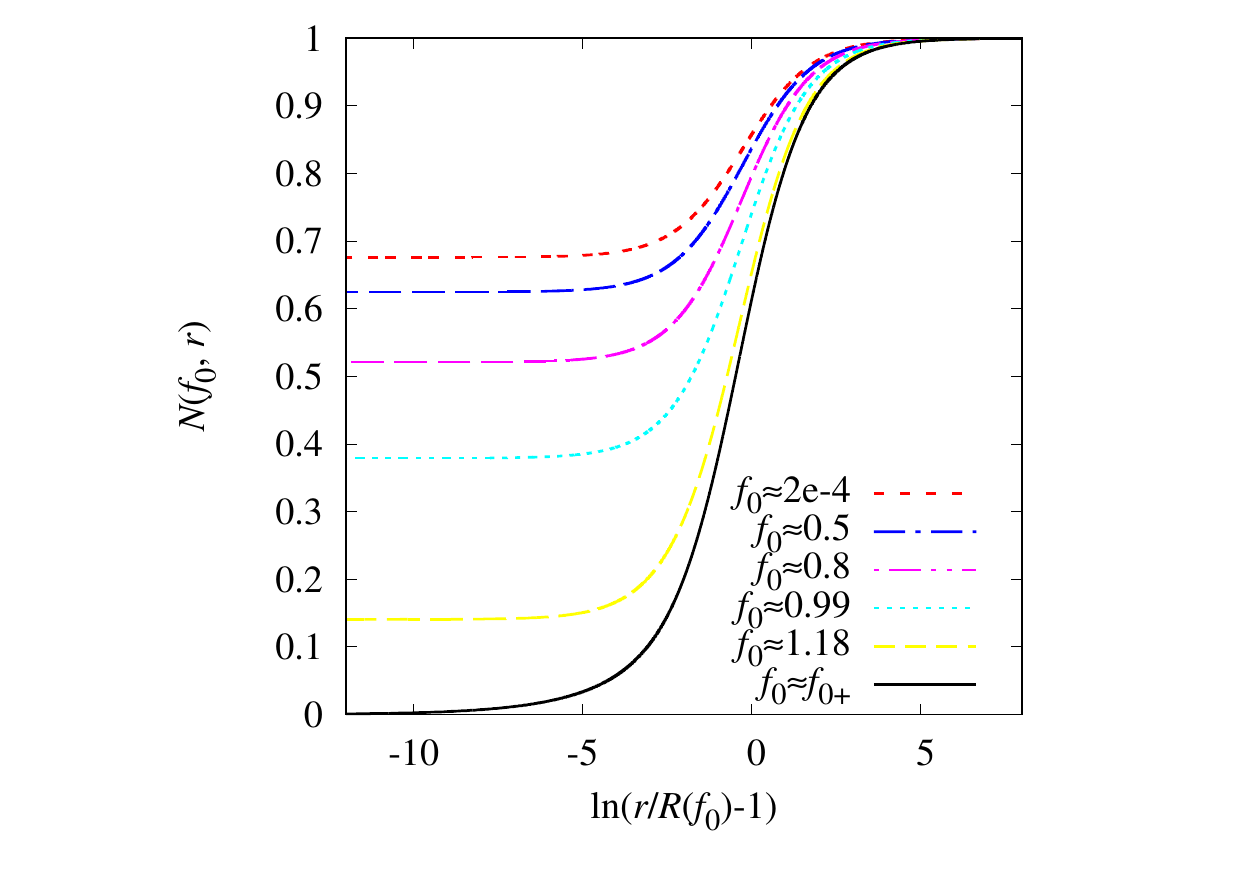}
\caption{The lapse function $N$ in the outer region for large
  $\chi_{s}$ ($\chi_{s}=0.8$), for several values of
  $f_{0}$.\label{fig:The-lapse-N-large_chi}}
\end{figure}

For large values of $\chi_{s}$ let us evaluate the lapse at
$f_0=f_{0,+}$ where $F(f_{0,+},R(f_{0,+}))=0$. Here we use time
$f_0(t) $ instead of $t$ to evaluate the lapse function.  At this
point, setting $\kappa_{0,+}=\kappa(f_{0,+})$, the lapse
$N(f_{0,+},R_{0+})$ is given by
\begin{align}
N(f_{0,+},R_{0+}) & =\left(1-\kappa_{,f_{0}}\dot{f}_{0}\int_{\infty}^{R_{0+}}\frac{1}{r_{1}^{2}}F\left(f_{0,+},r _{1}:\kappa_{0,+})\right)^{-3/2}\mathrm{d}r_{1}\right)\sqrt{F\left(f_{0,+},R_{0+}:\kappa_{0,+}\right)}
\nonumber \\
 & \approx\left(1-\frac{\kappa_{,f_{0}}\displaystyle{ {\int}_{0}^{1/R_{0+}}}
 F\left(f_{0,+},{1\over u}:\kappa_{0,+}\right)^{-3/2} du }{\kappa_{,f_{0}}\displaystyle{ {\int}_{0}^{1/R_{0+}}}
 F\left(f_{0,+},{1\over u}:\kappa_{0,+}\right)^{-3/2} du+\frac{[\kappa^{2}R_{\rm max}+R_{0+}^{4}(R_{\rm max}-r_{H})]R_{0+} R_{,f_{0}}}{\sqrt{R_{0+}^{4}-R_{0+}^{3}r_{H}+\kappa_{0,+}^{2}}\,[r_{H}\kappa_{0,+}\,(R_{\rm max}-R_{0+})]}}\right)
\nonumber  \\
 &
 ~~~\times \sqrt{F\left(f_{0,+},R_{0+}:\kappa_{0,+}\right)}=0\,,
\end{align}
because the quantity in parenthesis is of order unity, but $\sqrt{F}\to0$.
Along the same lines one can show that also for the internal metric
$N$ vanishes as $\dot{f}_{0}\to0$, so that the lapse, at $f_{0}=f_{0,+},$
vanishes up to $r=R_{0+}$. At this time we find that $\kappa_{1}<1$.
Therefore, for large $\chi_{s}$'s the end of collapse asymptotically
(in time-$t$) tends to a static metric having $0<\kappa_{1}<1$.
For other values of $r$ (and $f_{0}$), we need to evaluate
\begin{align}
 & N(f_{0},r) 
 =\sqrt{F(\kappa(f_{0}),r)}
\nonumber  
\\
 & \times \tiny{
\left(1-\frac{\kappa_{,f_{0}}\displaystyle{ {\int}_{0}^{1/r}}
 F\left(f_{0},{1\over u}\right)^{-3/2} du }{\kappa_{,f_{0}}
 \displaystyle{ {\int}_{0}^{1/R}}
 F\left(f_{0},{1\over u}\right)^{-3/2} du
 +\frac{[{\kappa^{2}\over R^4} R_{\rm max}+(R_{\rm max}-r_{H})]RR_{,f_{0}}}{\sqrt{ F\left(f_{0},R\right)}\,[\sqrt{ F\left(f_{0},R\right)}\,\sqrt{R}\sqrt{R_{\rm max}-r_{H}}\,\sqrt{r_{H}(R_{\rm max}-R)}+r_{H}{\kappa \over R^{2}}\,(R_{\rm max}-R)]}}\right)}
\,,
\end{align}
where $R=R\bigl(f(f_{0})\bigr)$ stands for the radius at the surface
of the star. We plot $N(f_{0+},r)$ as a function of $r$, in Fig.\ \ref{fig:The-lapse-N-large_chi}.

\section{Discussion and conclusions\label{sec:Discussion}}

The VCDM theory is a Type IIa minimally modified gravity (MMG). By
type IIa we mean a theory for which an Einstein frame does not exist,
i.e.\ we cannot rewrite the theory by means of the Einstein-Hilbert
action and matter fields coupled to gravity in a whatever non-trivial
way, and in which the propagation of gravitational waves is the same
as the propagation of electromagnetic waves~\cite{Aoki:2021zuy}.
Furthermore it is a MMG theory, which means that in the gravity
sector, the theory only has two local physical degrees of freedom, the
two polarizations of the standard tensorial gravitational waves. In
other words, this theory then does not add any additional degrees of
freedom in the gravity sector (as happens instead in standard
scalar-vector-tensor theories). The fact, the absence of extra degrees
of freedom in the gravity sector, means we do not need a mechanisms as
to screen them. On the other hand, one needs to see the solutions of
VCDM which can be tested against observations. For instance, one needs
to find VCDM solutions which describe black holes and stars. Such 
solutions, assuming spherical symmetry, are known to exist and they
coincide with GR solutions if both the trace of the extrinsic
curvature (for $t$-constant slicing) and the auxiliary field $\phi$
are constant~\cite{DeFelice:2020onz,DeFelice:2021xps}. The property
that VCDM and GR share common solutions was shown
in~\cite{DeFelice:2022uxv}, and it always holds in any spacetime
geometry provided that the auxiliary fields of VCDM, $\phi$ and
$\lambda$, are constant in time and space and that the time-$t$
foliation of the manifold admits a constant trace for the extrinsic
curvature, namely $K=K_{\infty}$. In this paper, we have shown that we
can successfully construct VCDM solutions for a spherically symmetric
collapse which are identical to the analogous solutions in GR. These
solutions consist of a spherically symmetric collapse endowed with a
foliation which keeps a constant extrinsic curvature during the collapse
itself.

In particular, as in the Oppenheimer-Snyder case, we have a cloud of dust with initial radius $R_{\rm max}$. For the inside (the dust cloud) metric, we have rewritten a closed homogeneous and isotropic metric in a coordinate system which has a time-$t$ slicing with $K=K_{\infty}$.  For the outer solution we have instead rewritten the standard Schwarzschild-de Sitter metric once so that its time-$t$ slicing allows $K=K_{\infty}$.  Then we use Israel junction conditions to find the appropriate matching conditions at the surface of the star. In addition, for VCDM we have to make sure that all the fields are smooth at the matching surface, in particular that the constant-$t$ hypersurface does not have any cusp at the junction surface. 

In a previous work,~\cite{DeFelice:2020onz}, it was shown that the stationary spherically symmetric solutions in VCDM are Schwarzschild solutions written in the following particular coordinate system
\begin{align}
ds^{2} & =-\frac{N^{2}}{F}\,[F-\beta^{2}]\,dt^{2}+2\,\frac{N\beta}{F}\,dt\,dr+\frac{dr^{2}}{F}+r^{2}\left[\frac{dz^{2}}{1-z^{2}}+(1-z^{2})\,d\theta_{2}^{2}\right]\,,\\
F & =1-\frac{r_{H}}{r}+\frac{\kappa_{0}^{2}}{r^{4}}\,,\\
\beta & =\frac{\kappa_{0}}{r^{2}}\,,\\
N & =\sqrt{F}\,,
\end{align}
where we have neglected the contribution coming from the effective
cosmological constant. We have a totally free real parameter
$\kappa_{0}$.  In this paper we have shown that on introducing
$\kappa_{1}=\kappa_{0}/\kappa_{+}$ with
$\kappa_{+}\equiv\frac{3}{16}\sqrt{3}r_{H}^{2}$, we see that for
$t\to\infty$, the collapsing time-dependent solution approaches the
static case solution with $0<\kappa_{1}<1$ (and mirror case for
negative $\kappa_{1}$'s).

More in detail, from the point of view of the (far-away-from-the-star)
cosmological observer, whose time corresponds to $t$, the surface of
the star always enters its own apparent horizon (located at $r=r_{H}$)
and keeps evolving until it reaches a configuration for which the
lapse $N\to 0$. However, reaching this point takes an infinite time
$t$. This point does not corresponds to the standard curvature
singularity of the Oppenheimer-Snyder solution. We have already
mentioned that this VCDM collapsing solution (i.e.\ not only the final
static case but also the time-dependent collapsing one) is also
present in GR; however, in GR, this behavior would corresponds merely
to the artifact of the coordinate/foliation choice. Instead in VCDM,
that breaks 4D-diffeomorphism, this foliation has physical meaning as
the foliation which is chosen by the shadowy mode present in the
theory. Different foliations correspond to intrinsically different
objects in VCDM, and not all foliations of a given GR metric
correspond to solutions of the VCDM equations of motion.

The presence of the vanishing lapse endpoint implies the necessity of
a UV completion to describe the physics inside the black hole beyond
this point. On the other hand, since the cosmic time $t$ at the
formation of this endpoint is infinite, VCDM can safely describe the
whole history of the universe at large scales without knowledge of the
unknown UV completion, despite the presence of the so-called shadowy
mode whose description requires proper boundary conditions. The same
final state could be a possible outcome for other theories which are
endowed with shadowy modes, and further investigations in this sense
could be interesting.

As stated above the final state of the solutions predicts that $0<\kappa_{1}<1$.  In particular, this parameter affects the exterior VCDM solutions, in particular it may influence the behavior of gravitational waves, and if so, it is actually possible to look for experimental bounds via the study of gravitational waves. We will defer such a study to a future project.  \selectlanguage{american}%

\begin{acknowledgments}
The work of A.D.F.\ was supported by Japan Society for the Promotion
of Science Grants-in-Aid for Scientific Research No.\ 20K03969. K.M.
would like to acknowledges the Yukawa Institute for Theoretical
Physics at Kyoto University, where the present work was proceeded
during the Visitors Program of FY2021 and FY2022.  The work of
K.M. was supported by JSPS KAKENHI Grant Numbers JP17H06359 and
JP19K03857. The work of S.M. is supported in part by Japan Society for
the Promotion of Science Grants-in-Aid for Scientific Research
No. 17H02890 and No. 17H06359 and by World Premier International
Research Center Initiative, the Ministry of Education, Culture,
Sports, Science and Technology, Japan. The work of M.C.P.\ is
supported by Mahidol University (Grand No. MD-PD\_MUMRC\_2022\_09) and
was supported by the Japan Society for the Promotion of Science
Grant-in-Aid for Scientific Research No.\ 17H06359 during the initial
phase of the project.
\end{acknowledgments}


\appendix

\section{Sturm theorem}

Let us study in a bit more details the function
\begin{equation}
F(t,r)=1-\frac{r_{H}}{r}+\frac{\kappa_{+}^{2}\kappa_{1}(t)^{2}}{r^{4}}\,,
\end{equation}
and in particular let us use the Sturm theorem as to determine the
number of real solutions of the equation $r^{4}F=0$. In particular,
let us study the zeros of $r^{4}F$ in the case $r\neq0$ and $\kappa_{1}\neq0$.
Then we have
\begin{equation}
r^{4}-r_{H}r^{3}+\kappa_{+}^{2}\kappa_{1}(t)^{2}=0\,.
\end{equation}
The discriminant of this polynomial is given by
\begin{equation}
\Delta=\frac{3^9 \kappa_{1}^{4}\left(\kappa_{1}^{2}-1\right)r_{H}^{12}}{2^{16}}\,.
\end{equation}
Then for $0<\kappa_{1}^{2}<1$ two solutions are real and two are
complex. For $\kappa_{1}^{2}=1$, there are (at least) two coincident
roots.

Let us now study the case $\kappa_{1}^{2}>1$. According to Sturm
theorem on univariate polynomials with real coefficients, one defines
\begin{align}
P_{0} & =r^{4}-r_{H}r^{3}+\kappa_{+}^{2}\kappa_{1}(t)^{2}\,,\\
P_{1} & =P'_{0}=4r^{3}-3r_{H}r^{2}\,,\\
P_{2} & =-{\rm rem}(P_{0},P_{1})=\frac{3}{16}r^{2}r_{H}^{2}-\frac{27}{256}\kappa_{1}^{2}r_{H}^{4}\,,\\
P_{3} & =-{\rm rem}(P_{1},P_{2})=-\frac{9}{4}\kappa_{1}^{2}rr_{H}^{2}+\frac{27}{16}\kappa_{1}^{2}r_{H}^{3}\,,\\
P_{4} & =-{\rm rem}(P_{2},P_{3})=\frac{27}{256}\left(\kappa_{1}^{2}-1\right)r_{H}^{4}\,.
\end{align}
Then the sign of these polynomials at $(-\infty,+\infty)$ is given
by $S_{1}=(+,-,+,+,+)$ and $S_{2}=(+,+,+,-,+)$. So the number of
real roots of $P_{0}$ is given by the difference between the number
$V$ of sign variations inside $S_{1,2}$, namely\footnote{For the case $0<\kappa_{1}<1$, we would have $(+,-,+,+,-)$ and $(+,+,+,-,-)$
as to have $V(-\infty)-V(+\infty)=3-1=2$ real roots.} $V(-\infty)-V(+\infty)=2-2=0$. So for $\kappa_{1}^{2}>1$ we have
no real roots.

\bibliographystyle{apsrev4-2}
\bibliography{bibliography}

\begin{thebibliography}{26}%
\makeatletter
\providecommand \@ifxundefined [1]{%
 \@ifx{#1\undefined}
}%
\providecommand \@ifnum [1]{%
 \ifnum #1\expandafter \@firstoftwo
 \else \expandafter \@secondoftwo
 \fi
}%
\providecommand \@ifx [1]{%
 \ifx #1\expandafter \@firstoftwo
 \else \expandafter \@secondoftwo
 \fi
}%
\providecommand \natexlab [1]{#1}%
\providecommand \enquote  [1]{``#1''}%
\providecommand \bibnamefont  [1]{#1}%
\providecommand \bibfnamefont [1]{#1}%
\providecommand \citenamefont [1]{#1}%
\providecommand \href@noop [0]{\@secondoftwo}%
\providecommand \href [0]{\begingroup \@sanitize@url \@href}%
\providecommand \@href[1]{\@@startlink{#1}\@@href}%
\providecommand \@@href[1]{\endgroup#1\@@endlink}%
\providecommand \@sanitize@url [0]{\catcode `\\12\catcode `\$12\catcode
  `\&12\catcode `\#12\catcode `\^12\catcode `\_12\catcode `\%12\relax}%
\providecommand \@@startlink[1]{}%
\providecommand \@@endlink[0]{}%
\providecommand \url  [0]{\begingroup\@sanitize@url \@url }%
\providecommand \@url [1]{\endgroup\@href {#1}{\urlprefix }}%
\providecommand \urlprefix  [0]{URL }%
\providecommand \Eprint [0]{\href }%
\providecommand \doibase [0]{https://doi.org/}%
\providecommand \selectlanguage [0]{\@gobble}%
\providecommand \bibinfo  [0]{\@secondoftwo}%
\providecommand \bibfield  [0]{\@secondoftwo}%
\providecommand \translation [1]{[#1]}%
\providecommand \BibitemOpen [0]{}%
\providecommand \bibitemStop [0]{}%
\providecommand \bibitemNoStop [0]{.\EOS\space}%
\providecommand \EOS [0]{\spacefactor3000\relax}%
\providecommand \BibitemShut  [1]{\csname bibitem#1\endcsname}%
\let\auto@bib@innerbib\@empty
\bibitem [{\citenamefont {De~Felice}\ \emph {et~al.}(2020)\citenamefont
  {De~Felice}, \citenamefont {Doll},\ and\ \citenamefont
  {Mukohyama}}]{DeFelice:2020eju}%
  \BibitemOpen
  \bibfield  {author} {\bibinfo {author} {\bibfnamefont {A.}~\bibnamefont
  {De~Felice}}, \bibinfo {author} {\bibfnamefont {A.}~\bibnamefont {Doll}},\
  and\ \bibinfo {author} {\bibfnamefont {S.}~\bibnamefont {Mukohyama}},\ }\href
  {https://doi.org/10.1088/1475-7516/2020/09/034} {\bibfield  {journal}
  {\bibinfo  {journal} {JCAP}\ }\textbf {\bibinfo {volume} {09}},\ \bibinfo
  {pages} {034}},\ \Eprint {https://arxiv.org/abs/2004.12549} {arXiv:2004.12549
  [gr-qc]} \BibitemShut {NoStop}%
\bibitem [{\citenamefont {De~Felice}\ and\ \citenamefont
  {Mukohyama}(2021)}]{DeFelice:2020prd}%
  \BibitemOpen
  \bibfield  {author} {\bibinfo {author} {\bibfnamefont {A.}~\bibnamefont
  {De~Felice}}\ and\ \bibinfo {author} {\bibfnamefont {S.}~\bibnamefont
  {Mukohyama}},\ }\href {https://doi.org/10.1088/1475-7516/2021/04/018}
  {\bibfield  {journal} {\bibinfo  {journal} {JCAP}\ }\textbf {\bibinfo
  {volume} {04}},\ \bibinfo {pages} {018}},\ \Eprint
  {https://arxiv.org/abs/2011.04188} {arXiv:2011.04188 [astro-ph.CO]}
  \BibitemShut {NoStop}%
\bibitem [{\citenamefont {De~Felice}\ \emph
  {et~al.}(2022{\natexlab{a}})\citenamefont {De~Felice}, \citenamefont {Maeda},
  \citenamefont {Mukohyama},\ and\ \citenamefont
  {Pookkillath}}]{DeFelice:2022uxv}%
  \BibitemOpen
  \bibfield  {author} {\bibinfo {author} {\bibfnamefont {A.}~\bibnamefont
  {De~Felice}}, \bibinfo {author} {\bibfnamefont {K.-i.}\ \bibnamefont
  {Maeda}}, \bibinfo {author} {\bibfnamefont {S.}~\bibnamefont {Mukohyama}},\
  and\ \bibinfo {author} {\bibfnamefont {M.~C.}\ \bibnamefont {Pookkillath}},\
  }\href {https://doi.org/10.1103/PhysRevD.106.024028} {\bibfield  {journal}
  {\bibinfo  {journal} {Phys. Rev. D}\ }\textbf {\bibinfo {volume} {106}},\
  \bibinfo {pages} {024028} (\bibinfo {year} {2022}{\natexlab{a}})},\ \Eprint
  {https://arxiv.org/abs/2204.08294} {arXiv:2204.08294 [gr-qc]} \BibitemShut
  {NoStop}%
\bibitem [{\citenamefont {De~Felice}\ \emph
  {et~al.}(2022{\natexlab{b}})\citenamefont {De~Felice}, \citenamefont
  {Mukohyama},\ and\ \citenamefont {Pookkillath}}]{DeFelice:2021xps}%
  \BibitemOpen
  \bibfield  {author} {\bibinfo {author} {\bibfnamefont {A.}~\bibnamefont
  {De~Felice}}, \bibinfo {author} {\bibfnamefont {S.}~\bibnamefont
  {Mukohyama}},\ and\ \bibinfo {author} {\bibfnamefont {M.~C.}\ \bibnamefont
  {Pookkillath}},\ }\href {https://doi.org/10.1103/PhysRevD.105.104013}
  {\bibfield  {journal} {\bibinfo  {journal} {Phys. Rev. D}\ }\textbf {\bibinfo
  {volume} {105}},\ \bibinfo {pages} {104013} (\bibinfo {year}
  {2022}{\natexlab{b}})},\ \Eprint {https://arxiv.org/abs/2110.14496}
  {arXiv:2110.14496 [gr-qc]} \BibitemShut {NoStop}%
\bibitem [{\citenamefont {De~Felice}\ \emph
  {et~al.}(2021{\natexlab{a}})\citenamefont {De~Felice}, \citenamefont
  {Mukohyama},\ and\ \citenamefont {Pookkillath}}]{DeFelice:2020cpt}%
  \BibitemOpen
  \bibfield  {author} {\bibinfo {author} {\bibfnamefont {A.}~\bibnamefont
  {De~Felice}}, \bibinfo {author} {\bibfnamefont {S.}~\bibnamefont
  {Mukohyama}},\ and\ \bibinfo {author} {\bibfnamefont {M.~C.}\ \bibnamefont
  {Pookkillath}},\ }\href {https://doi.org/10.1016/j.physletb.2021.136201}
  {\bibfield  {journal} {\bibinfo  {journal} {Phys. Lett. B}\ }\textbf
  {\bibinfo {volume} {816}},\ \bibinfo {pages} {136201} (\bibinfo {year}
  {2021}{\natexlab{a}})},\ \Eprint {https://arxiv.org/abs/2009.08718}
  {arXiv:2009.08718 [astro-ph.CO]} \BibitemShut {NoStop}%
\bibitem [{\citenamefont {De~Felice}\ \emph
  {et~al.}(2021{\natexlab{b}})\citenamefont {De~Felice}, \citenamefont {Doll},
  \citenamefont {Larrouturou},\ and\ \citenamefont
  {Mukohyama}}]{DeFelice:2020onz}%
  \BibitemOpen
  \bibfield  {author} {\bibinfo {author} {\bibfnamefont {A.}~\bibnamefont
  {De~Felice}}, \bibinfo {author} {\bibfnamefont {A.}~\bibnamefont {Doll}},
  \bibinfo {author} {\bibfnamefont {F.}~\bibnamefont {Larrouturou}},\ and\
  \bibinfo {author} {\bibfnamefont {S.}~\bibnamefont {Mukohyama}},\ }\href
  {https://doi.org/10.1088/1475-7516/2021/03/004} {\bibfield  {journal}
  {\bibinfo  {journal} {JCAP}\ }\textbf {\bibinfo {volume} {03}},\ \bibinfo
  {pages} {004}},\ \Eprint {https://arxiv.org/abs/2010.13067} {arXiv:2010.13067
  [gr-qc]} \BibitemShut {NoStop}%
\bibitem [{\citenamefont {Aoki}\ \emph {et~al.}(2021)\citenamefont {Aoki},
  \citenamefont {Di~Filippo},\ and\ \citenamefont {Mukohyama}}]{Aoki:2021zuy}%
  \BibitemOpen
  \bibfield  {author} {\bibinfo {author} {\bibfnamefont {K.}~\bibnamefont
  {Aoki}}, \bibinfo {author} {\bibfnamefont {F.}~\bibnamefont {Di~Filippo}},\
  and\ \bibinfo {author} {\bibfnamefont {S.}~\bibnamefont {Mukohyama}},\ }\href
  {https://doi.org/10.1088/1475-7516/2021/05/071} {\bibfield  {journal}
  {\bibinfo  {journal} {JCAP}\ }\textbf {\bibinfo {volume} {05}},\ \bibinfo
  {pages} {071}},\ \Eprint {https://arxiv.org/abs/2103.15044} {arXiv:2103.15044
  [gr-qc]} \BibitemShut {NoStop}%
\bibitem [{\citenamefont {Afshordi}\ \emph {et~al.}(2007)\citenamefont
  {Afshordi}, \citenamefont {Chung},\ and\ \citenamefont
  {Geshnizjani}}]{Afshordi:2006ad}%
  \BibitemOpen
  \bibfield  {author} {\bibinfo {author} {\bibfnamefont {N.}~\bibnamefont
  {Afshordi}}, \bibinfo {author} {\bibfnamefont {D.~J.~H.}\ \bibnamefont
  {Chung}},\ and\ \bibinfo {author} {\bibfnamefont {G.}~\bibnamefont
  {Geshnizjani}},\ }\href {https://doi.org/10.1103/PhysRevD.75.083513}
  {\bibfield  {journal} {\bibinfo  {journal} {Phys. Rev. D}\ }\textbf {\bibinfo
  {volume} {75}},\ \bibinfo {pages} {083513} (\bibinfo {year} {2007})},\
  \Eprint {https://arxiv.org/abs/hep-th/0609150} {arXiv:hep-th/0609150}
  \BibitemShut {NoStop}%
\bibitem [{\citenamefont {Maeda}\ and\ \citenamefont
  {Panpanich}(2022)}]{Maeda:2022ozc}%
  \BibitemOpen
  \bibfield  {author} {\bibinfo {author} {\bibfnamefont {K.-i.}\ \bibnamefont
  {Maeda}}\ and\ \bibinfo {author} {\bibfnamefont {S.}~\bibnamefont
  {Panpanich}},\ }\href@noop {} {\  (\bibinfo {year} {2022})},\ \Eprint
  {https://arxiv.org/abs/2202.04908} {arXiv:2202.04908 [gr-qc]} \BibitemShut
  {NoStop}%
\bibitem [{\citenamefont {Kohri}\ and\ \citenamefont
  {Maeda}(2022)}]{Kohri:2022vst}%
  \BibitemOpen
  \bibfield  {author} {\bibinfo {author} {\bibfnamefont {K.}~\bibnamefont
  {Kohri}}\ and\ \bibinfo {author} {\bibfnamefont {K.-i.}\ \bibnamefont
  {Maeda}},\ }\href {https://doi.org/10.1093/ptep/ptac114} {\bibfield
  {journal} {\bibinfo  {journal} {PTEP}\ }\textbf {\bibinfo {volume} {2022}},\
  \bibinfo {pages} {091E01} (\bibinfo {year} {2022})},\ \Eprint
  {https://arxiv.org/abs/2206.11257} {arXiv:2206.11257 [gr-qc]} \BibitemShut
  {NoStop}%
\bibitem [{\citenamefont {De~Felice}\ and\ \citenamefont
  {Mukohyama}(2016{\natexlab{a}})}]{DeFelice:2015hla}%
  \BibitemOpen
  \bibfield  {author} {\bibinfo {author} {\bibfnamefont {A.}~\bibnamefont
  {De~Felice}}\ and\ \bibinfo {author} {\bibfnamefont {S.}~\bibnamefont
  {Mukohyama}},\ }\href {https://doi.org/10.1016/j.physletb.2015.11.050}
  {\bibfield  {journal} {\bibinfo  {journal} {Phys. Lett. B}\ }\textbf
  {\bibinfo {volume} {752}},\ \bibinfo {pages} {302} (\bibinfo {year}
  {2016}{\natexlab{a}})},\ \Eprint {https://arxiv.org/abs/1506.01594}
  {arXiv:1506.01594 [hep-th]} \BibitemShut {NoStop}%
\bibitem [{\citenamefont {De~Felice}\ and\ \citenamefont
  {Mukohyama}(2016{\natexlab{b}})}]{DeFelice:2015moy}%
  \BibitemOpen
  \bibfield  {author} {\bibinfo {author} {\bibfnamefont {A.}~\bibnamefont
  {De~Felice}}\ and\ \bibinfo {author} {\bibfnamefont {S.}~\bibnamefont
  {Mukohyama}},\ }\href {https://doi.org/10.1088/1475-7516/2016/04/028}
  {\bibfield  {journal} {\bibinfo  {journal} {JCAP}\ }\textbf {\bibinfo
  {volume} {04}},\ \bibinfo {pages} {028}},\ \Eprint
  {https://arxiv.org/abs/1512.04008} {arXiv:1512.04008 [hep-th]} \BibitemShut
  {NoStop}%
\bibitem [{\citenamefont {De~Felice}\ \emph
  {et~al.}(2021{\natexlab{c}})\citenamefont {De~Felice}, \citenamefont
  {Mukohyama},\ and\ \citenamefont {Pookkillath}}]{DeFelice:2021trp}%
  \BibitemOpen
  \bibfield  {author} {\bibinfo {author} {\bibfnamefont {A.}~\bibnamefont
  {De~Felice}}, \bibinfo {author} {\bibfnamefont {S.}~\bibnamefont
  {Mukohyama}},\ and\ \bibinfo {author} {\bibfnamefont {M.~C.}\ \bibnamefont
  {Pookkillath}},\ }\href {https://doi.org/10.1088/1475-7516/2021/12/011}
  {\bibfield  {journal} {\bibinfo  {journal} {JCAP}\ }\textbf {\bibinfo
  {volume} {12}}\bibfield  {number} {\bibinfo  {number} { (12)},\ \bibinfo
  {pages} {011}},\ }\Eprint {https://arxiv.org/abs/2110.01237}
  {arXiv:2110.01237 [astro-ph.CO]} \BibitemShut {NoStop}%
\bibitem [{\citenamefont {De~Felice}\ \emph
  {et~al.}(2017{\natexlab{a}})\citenamefont {De~Felice}, \citenamefont
  {Mukohyama},\ and\ \citenamefont {Oliosi}}]{DeFelice:2017wel}%
  \BibitemOpen
  \bibfield  {author} {\bibinfo {author} {\bibfnamefont {A.}~\bibnamefont
  {De~Felice}}, \bibinfo {author} {\bibfnamefont {S.}~\bibnamefont
  {Mukohyama}},\ and\ \bibinfo {author} {\bibfnamefont {M.}~\bibnamefont
  {Oliosi}},\ }\href {https://doi.org/10.1103/PhysRevD.96.024032} {\bibfield
  {journal} {\bibinfo  {journal} {Phys. Rev. D}\ }\textbf {\bibinfo {volume}
  {96}},\ \bibinfo {pages} {024032} (\bibinfo {year} {2017}{\natexlab{a}})},\
  \Eprint {https://arxiv.org/abs/1701.01581} {arXiv:1701.01581 [hep-th]}
  \BibitemShut {NoStop}%
\bibitem [{\citenamefont {De~Felice}\ \emph
  {et~al.}(2017{\natexlab{b}})\citenamefont {De~Felice}, \citenamefont
  {Mukohyama},\ and\ \citenamefont {Oliosi}}]{DeFelice:2017rli}%
  \BibitemOpen
  \bibfield  {author} {\bibinfo {author} {\bibfnamefont {A.}~\bibnamefont
  {De~Felice}}, \bibinfo {author} {\bibfnamefont {S.}~\bibnamefont
  {Mukohyama}},\ and\ \bibinfo {author} {\bibfnamefont {M.}~\bibnamefont
  {Oliosi}},\ }\href {https://doi.org/10.1103/PhysRevD.96.104036} {\bibfield
  {journal} {\bibinfo  {journal} {Phys. Rev. D}\ }\textbf {\bibinfo {volume}
  {96}},\ \bibinfo {pages} {104036} (\bibinfo {year} {2017}{\natexlab{b}})},\
  \Eprint {https://arxiv.org/abs/1709.03108} {arXiv:1709.03108 [hep-th]}
  \BibitemShut {NoStop}%
\bibitem [{\citenamefont {Mukohyama}\ and\ \citenamefont
  {Noui}(2019)}]{Mukohyama:2019unx}%
  \BibitemOpen
  \bibfield  {author} {\bibinfo {author} {\bibfnamefont {S.}~\bibnamefont
  {Mukohyama}}\ and\ \bibinfo {author} {\bibfnamefont {K.}~\bibnamefont
  {Noui}},\ }\href {https://doi.org/10.1088/1475-7516/2019/07/049} {\bibfield
  {journal} {\bibinfo  {journal} {JCAP}\ }\textbf {\bibinfo {volume} {07}},\
  \bibinfo {pages} {049}},\ \Eprint {https://arxiv.org/abs/1905.02000}
  {arXiv:1905.02000 [gr-qc]} \BibitemShut {NoStop}%
\bibitem [{\citenamefont {Aoki}\ \emph
  {et~al.}(2020{\natexlab{a}})\citenamefont {Aoki}, \citenamefont {De~Felice},
  \citenamefont {Mukohyama}, \citenamefont {Noui}, \citenamefont {Oliosi},\
  and\ \citenamefont {Pookkillath}}]{Aoki:2020oqc}%
  \BibitemOpen
  \bibfield  {author} {\bibinfo {author} {\bibfnamefont {K.}~\bibnamefont
  {Aoki}}, \bibinfo {author} {\bibfnamefont {A.}~\bibnamefont {De~Felice}},
  \bibinfo {author} {\bibfnamefont {S.}~\bibnamefont {Mukohyama}}, \bibinfo
  {author} {\bibfnamefont {K.}~\bibnamefont {Noui}}, \bibinfo {author}
  {\bibfnamefont {M.}~\bibnamefont {Oliosi}},\ and\ \bibinfo {author}
  {\bibfnamefont {M.~C.}\ \bibnamefont {Pookkillath}},\ }\href
  {https://doi.org/10.1140/epjc/s10052-020-8291-1} {\bibfield  {journal}
  {\bibinfo  {journal} {Eur. Phys. J. C}\ }\textbf {\bibinfo {volume} {80}},\
  \bibinfo {pages} {708} (\bibinfo {year} {2020}{\natexlab{a}})},\ \Eprint
  {https://arxiv.org/abs/2005.13972} {arXiv:2005.13972 [astro-ph.CO]}
  \BibitemShut {NoStop}%
\bibitem [{\citenamefont {Aoki}\ \emph
  {et~al.}(2020{\natexlab{b}})\citenamefont {Aoki}, \citenamefont {Gorji},\
  and\ \citenamefont {Mukohyama}}]{Aoki:2020lig}%
  \BibitemOpen
  \bibfield  {author} {\bibinfo {author} {\bibfnamefont {K.}~\bibnamefont
  {Aoki}}, \bibinfo {author} {\bibfnamefont {M.~A.}\ \bibnamefont {Gorji}},\
  and\ \bibinfo {author} {\bibfnamefont {S.}~\bibnamefont {Mukohyama}},\ }\href
  {https://doi.org/10.1016/j.physletb.2020.135843} {\bibfield  {journal}
  {\bibinfo  {journal} {Phys. Lett. B}\ }\textbf {\bibinfo {volume} {810}},\
  \bibinfo {pages} {135843} (\bibinfo {year} {2020}{\natexlab{b}})},\ \Eprint
  {https://arxiv.org/abs/2005.03859} {arXiv:2005.03859 [gr-qc]} \BibitemShut
  {NoStop}%
\bibitem [{\citenamefont {De~Felice}\ \emph
  {et~al.}(2021{\natexlab{d}})\citenamefont {De~Felice}, \citenamefont
  {Larrouturou}, \citenamefont {Mukohyama},\ and\ \citenamefont
  {Oliosi}}]{DeFelice:2020ecp}%
  \BibitemOpen
  \bibfield  {author} {\bibinfo {author} {\bibfnamefont {A.}~\bibnamefont
  {De~Felice}}, \bibinfo {author} {\bibfnamefont {F.}~\bibnamefont
  {Larrouturou}}, \bibinfo {author} {\bibfnamefont {S.}~\bibnamefont
  {Mukohyama}},\ and\ \bibinfo {author} {\bibfnamefont {M.}~\bibnamefont
  {Oliosi}},\ }\href {https://doi.org/10.1088/1475-7516/2021/04/015} {\bibfield
   {journal} {\bibinfo  {journal} {JCAP}\ }\textbf {\bibinfo {volume} {04}},\
  \bibinfo {pages} {015}},\ \Eprint {https://arxiv.org/abs/2012.01073}
  {arXiv:2012.01073 [gr-qc]} \BibitemShut {NoStop}%
\bibitem [{\citenamefont {De~Felice}\ \emph
  {et~al.}(2022{\natexlab{c}})\citenamefont {De~Felice}, \citenamefont
  {Mukohyama},\ and\ \citenamefont {Pookkillath}}]{DeFelice:2022mcd}%
  \BibitemOpen
  \bibfield  {author} {\bibinfo {author} {\bibfnamefont {A.}~\bibnamefont
  {De~Felice}}, \bibinfo {author} {\bibfnamefont {S.}~\bibnamefont
  {Mukohyama}},\ and\ \bibinfo {author} {\bibfnamefont {M.~C.}\ \bibnamefont
  {Pookkillath}},\ }\href {https://doi.org/10.1103/PhysRevD.106.084050}
  {\bibfield  {journal} {\bibinfo  {journal} {Phys. Rev. D}\ }\textbf {\bibinfo
  {volume} {106}},\ \bibinfo {pages} {084050} (\bibinfo {year}
  {2022}{\natexlab{c}})},\ \Eprint {https://arxiv.org/abs/2206.03338}
  {arXiv:2206.03338 [gr-qc]} \BibitemShut {NoStop}%
\bibitem [{\citenamefont {De~Felice}\ \emph {et~al.}(2018)\citenamefont
  {De~Felice}, \citenamefont {Larrouturou}, \citenamefont {Mukohyama},\ and\
  \citenamefont {Oliosi}}]{DeFelice:2018vza}%
  \BibitemOpen
  \bibfield  {author} {\bibinfo {author} {\bibfnamefont {A.}~\bibnamefont
  {De~Felice}}, \bibinfo {author} {\bibfnamefont {F.}~\bibnamefont
  {Larrouturou}}, \bibinfo {author} {\bibfnamefont {S.}~\bibnamefont
  {Mukohyama}},\ and\ \bibinfo {author} {\bibfnamefont {M.}~\bibnamefont
  {Oliosi}},\ }\href {https://doi.org/10.1103/PhysRevD.98.104031} {\bibfield
  {journal} {\bibinfo  {journal} {Phys. Rev. D}\ }\textbf {\bibinfo {volume}
  {98}},\ \bibinfo {pages} {104031} (\bibinfo {year} {2018})},\ \Eprint
  {https://arxiv.org/abs/1808.01403} {arXiv:1808.01403 [gr-qc]} \BibitemShut
  {NoStop}%
\bibitem [{\citenamefont {Estabrook}\ \emph {et~al.}(1973)\citenamefont
  {Estabrook}, \citenamefont {Wahlquist}, \citenamefont {Christensen},
  \citenamefont {DeWitt}, \citenamefont {Smarr},\ and\ \citenamefont
  {Tsiang}}]{Estabrook:1973ue}%
  \BibitemOpen
  \bibfield  {author} {\bibinfo {author} {\bibfnamefont {F.}~\bibnamefont
  {Estabrook}}, \bibinfo {author} {\bibfnamefont {H.}~\bibnamefont
  {Wahlquist}}, \bibinfo {author} {\bibfnamefont {S.}~\bibnamefont
  {Christensen}}, \bibinfo {author} {\bibfnamefont {B.}~\bibnamefont {DeWitt}},
  \bibinfo {author} {\bibfnamefont {L.}~\bibnamefont {Smarr}},\ and\ \bibinfo
  {author} {\bibfnamefont {E.}~\bibnamefont {Tsiang}},\ }\href
  {https://doi.org/10.1103/PhysRevD.7.2814} {\bibfield  {journal} {\bibinfo
  {journal} {Phys. Rev. D}\ }\textbf {\bibinfo {volume} {7}},\ \bibinfo {pages}
  {2814} (\bibinfo {year} {1973})}\BibitemShut {NoStop}%
\bibitem [{\citenamefont {Eardley}\ and\ \citenamefont
  {Smarr}(1979)}]{Eardley:1979}%
  \BibitemOpen
  \bibfield  {author} {\bibinfo {author} {\bibfnamefont {D.~M.}\ \bibnamefont
  {Eardley}}\ and\ \bibinfo {author} {\bibfnamefont {L.}~\bibnamefont
  {Smarr}},\ }\href {https://doi.org/10.1103/PhysRevD.19.2239} {\bibfield
  {journal} {\bibinfo  {journal} {Phys. Rev. D}\ }\textbf {\bibinfo {volume}
  {19}},\ \bibinfo {pages} {2239} (\bibinfo {year} {1979})}\BibitemShut
  {NoStop}%
\bibitem [{\citenamefont {Maeda}(1980)}]{Maeda:1980}%
  \BibitemOpen
  \bibfield  {author} {\bibinfo {author} {\bibfnamefont {K.-i.}\ \bibnamefont
  {Maeda}},\ }\href {https://doi.org/10.1143/PTP.63.425} {\bibfield  {journal}
  {\bibinfo  {journal} {Progress of Theoretical Physics}\ }\textbf {\bibinfo
  {volume} {63}},\ \bibinfo {pages} {425} (\bibinfo {year} {1980})},\ \Eprint
  {https://arxiv.org/abs/https://academic.oup.com/ptp/article-pdf/63/2/425/5333183/63-2-425.pdf}
  {https://academic.oup.com/ptp/article-pdf/63/2/425/5333183/63-2-425.pdf}
  \BibitemShut {NoStop}%
\bibitem [{\citenamefont {Petrich}\ \emph {et~al.}(1985)\citenamefont
  {Petrich}, \citenamefont {Shapiro},\ and\ \citenamefont
  {Teukolsky}}]{Petrich:1985}%
  \BibitemOpen
  \bibfield  {author} {\bibinfo {author} {\bibfnamefont {L.~I.}\ \bibnamefont
  {Petrich}}, \bibinfo {author} {\bibfnamefont {S.~L.}\ \bibnamefont
  {Shapiro}},\ and\ \bibinfo {author} {\bibfnamefont {S.~A.}\ \bibnamefont
  {Teukolsky}},\ }\href {https://doi.org/10.1103/PhysRevD.31.2459} {\bibfield
  {journal} {\bibinfo  {journal} {Phys. Rev. D}\ }\textbf {\bibinfo {volume}
  {31}},\ \bibinfo {pages} {2459} (\bibinfo {year} {1985})}\BibitemShut
  {NoStop}%
\bibitem [{\citenamefont {Nakao}\ \emph {et~al.}(1991)\citenamefont {Nakao},
  \citenamefont {Maeda}, \citenamefont {Nakamura},\ and\ \citenamefont
  {Oohara}}]{Nakao:1990gw}%
  \BibitemOpen
  \bibfield  {author} {\bibinfo {author} {\bibfnamefont {K.-I.}\ \bibnamefont
  {Nakao}}, \bibinfo {author} {\bibfnamefont {K.-I.}\ \bibnamefont {Maeda}},
  \bibinfo {author} {\bibfnamefont {T.}~\bibnamefont {Nakamura}},\ and\
  \bibinfo {author} {\bibfnamefont {K.-I.}\ \bibnamefont {Oohara}},\ }\href
  {https://doi.org/10.1103/PhysRevD.44.1326} {\bibfield  {journal} {\bibinfo
  {journal} {Phys. Rev. D}\ }\textbf {\bibinfo {volume} {44}},\ \bibinfo
  {pages} {1326} (\bibinfo {year} {1991})}\BibitemShut {NoStop}%
\end{thebibliography}%

\end{document}